\documentclass{article}

\usepackage{arxiv}
\usepackage{enumitem}
\usepackage[utf8]{inputenc} 
\usepackage[T1]{fontenc}    
\usepackage{hyperref}       
\usepackage{url}            
\usepackage{booktabs}
\newcommand{\botrule}{\bottomrule}
\usepackage{amsfonts}       
\usepackage{nicefrac}       
\usepackage{microtype}      
\usepackage{lipsum}		
\usepackage{graphicx}
\usepackage{natbib}
\usepackage{doi}
\usepackage{tabularx}
\usepackage{amsmath}

\title{ELISA: An Interpretable Hybrid Generative AI Agent for Expression-Grounded Discovery in Single-Cell Genomics}

\author{
  Omar Coser\thanks{Correspondence author: Omar Coser. This manuscript has been submitted for peer review.} \\
  \texttt{omarcoser10@gmail.com} \\
}

\renewcommand{\shorttitle}{\textit{arXiv} Template}

\hypersetup{
pdftitle={A template for the arxiv style},
pdfsubject={q-bio.NC, q-bio.QM},
pdfauthor={David S.~Hippocampus, Elias D.~Striatum},
pdfkeywords={First keyword, Second keyword, More},
}

\begin{document}
\maketitle

\begin{abstract}
Translating single-cell RNA sequencing (scRNA-seq) data into mechanistic biological hypotheses remains a critical bottleneck, as agentic AI systems lack direct access to transcriptomic representations while expression foundation models remain opaque to natural language. Here we introduce ELISA (Embedding-Linked Interactive Single-cell Agent), an interpretable framework that unifies scGPT expression embeddings with BioBERT-based semantic retrieval and LLM-mediated interpretation for interactive single-cell discovery. An automatic query classifier routes inputs to gene marker scoring, semantic matching, or reciprocal rank fusion pipelines depending on whether the query is a gene signature, natural language concept, or mixture of both. Integrated analytical modules perform pathway activity scoringacross 60+ gene sets, ligand-receptor interaction prediction using 280+ curated pairs, condition-aware comparative analysis, and cell-type proportion estimation all operating directly on embedded data without access to the original count matrix. Benchmarked across six diverse scRNA-seq datasets spanning inflammatory lung
disease, pediatric and adult cancers, organoid models, healthy tissue, and
neurodevelopment, ELISA significantly outperforms CellWhisperer, a
classical lexical retriever (BM25), and a random baseline in cell type retrieval
(combined permutation test, $p < 2\times10^{-5}$ for each), with particularly
large gains on gene-signature queries (Cohen's $d = 5.98$ for MRR). ELISA
replicates published biological findings (mean composite score 0.88), and generates candidate hypotheses through grounded LLM reasoning,
bridging the gap between transcriptomic data exploration and biological
discovery.
\end{abstract}


\section{Introduction}
Single-cell RNA sequencing (scRNA-seq) has transformed our understanding of cellular heterogeneity by enabling genome-wide transcriptional profiling at single-cell resolution~\cite{tang2009mrna}. Standardized analytical pipelines support quality control, normalization, clustering, differential expression, and trajectory inference~\cite{luecken2019current}, catalyzing the construction of comprehensive cell atlases across tissues, developmental stages, and disease contexts. However a critical bottleneck persists: translating statistical outputs of differentially expressed gene lists, enriched pathways, and predicted ligand receptor interactions into mechanistic biological hypotheses remains labor-intensive, context-dependent, and difficult to scale or reproduce.

Large-language models (LLMs) offer a potential solution to this problem. LLMs encode substantial biomedical knowledge and perform competitively on clinical reasoning benchmarks~\cite{singhal2023large}, whereas retrieval-augmented generation (RAG) improves factual accuracy by grounding outputs in external knowledge at inference time~\cite{lewis2020retrieval}. These capabilities have motivated \emph{agentic} AI architectures that are capable of autonomous planning, tool usage, and iterative reasoning within closed-loop workflows.

Recent agentic systems span a broad range of biomedical applications (Table~\ref{tab:agents_comparison}). \textbf{Towards an AI Co-Scientist}~\cite{gottweis2025towards} introduces multi-agent hypothesis generation through structured debate and evolutionary refinement, though it operates over textual knowledge without interfacing with experimental data. \textbf{Biomni}~\cite{huang2025biomni} constructs a unified action space from biomedical tools and databases, enabling dynamic task orchestration including gene prioritization. \textbf{GeneAgent}~\cite{wang2025geneagent} and related systems~\cite{gao2024empowering} extend LLM reasoning to gene-set analysis, whereas \textbf{Virtual Lab}~\cite{swanson2025virtual} demonstrates collaborative multi-agent discovery. Within single-cell analysis, \textbf{CellAgent}~\cite{xiao2024cellagent} decomposes scRNA-seq workflows into agent-handled subtasks, \textbf{AutoBA}~\cite{zhou2023automated} generates executable pipelines from natural language, and \textbf{BRAD}~\cite{pickard2025automatic} integrates LLMs with enrichment analysis for biomarker identification. In retrieval-augmented space, \textbf{GeneGPT}~\cite{jin2024genegpt} provides structured access to NCBI databases, and systems for deep phenotyping~\cite{garcia2025improving} and biomedical data extraction~\cite{cinquin2024chip,niyonkuru2025leveraging} have demonstrated the utility of RAG for factual grounding. \textbf{CRISPR-GPT}~\cite{qu2025crispr} further illustrates agentic automation for gene-editing experiment design. However, these systems are primarily responsible for curated text and structured databases and lack the capacity to operate directly on high-dimensional transcriptomic representations.

Concurrently, foundation models for single-cell biology have achieved remarkable progress in the learning of expressive latent representations from transcriptomic data. \textbf{scGPT}~\cite{cui2024scgpt} employs generative pre-training over millions of single-cell transcriptomes, capturing gene-gene dependencies for cell embedding, annotation transfer, and perturbation prediction. Extensions such as \textbf{scWGBS-GPT}~\cite{liang2025scwgbs} and \textbf{Tokensome}~\cite{zhang2024tokensome} broaden learned representations to methylomics and multimodal settings. However, these expression embeddings are not designed for semantic querying; they capture transcriptional similarity in latent spaces that lack alignment with the natural language concepts that biologists use to formulate hypotheses. Notably, the \textbf{CellWhisperer}~\cite{schaefer2025multimodal} addressed part of this gap by learning joint embeddings of transcriptomes and textual annotations via contrastive training, enabling chat-based interrogation of scRNA-seq data within CELLxGENE~\cite{schaefer2025multimodal}. While this establishes a compelling proof of concept for natural-language exploration, it does not incorporate built-in analytical modules for pathway scoring, interaction prediction, or condition-aware comparison.

\begin{table*}[t]
\centering
\caption{Comparison of existing AI systems for biomedical and single-cell analysis. \textbf{Expr. Emb.}: uses expression-derived embeddings from foundation models; \textbf{Sem. Ret.}: semantic retrieval over biological annotations; \textbf{L--R / Pathway}: ligand--receptor interaction and pathway scoring from data; \textbf{Cond. Comp.}: condition-aware comparative analysis; \textbf{Interp. Report}: automated interpretive report generation with LLM.}
\label{tab:agents_comparison}
\small
\begin{tabular}{lcccccl}
\hline
\textbf{System} & \textbf{Expr.} & \textbf{Sem.} & \textbf{L--R /} & \textbf{Cond.} & \textbf{Interp.} & \textbf{Primary} \\
 & \textbf{Emb.} & \textbf{Ret.} & \textbf{Pathway} & \textbf{Comp.} & \textbf{Report} & \textbf{Scope} \\
\hline
AI Co-Scientist~\cite{gottweis2025towards} & -- & -- & -- & -- & \checkmark & Hypothesis generation \\
Biomni~\cite{huang2025biomni} & -- & \checkmark & -- & -- & -- & General biomedical \\
GeneAgent~\cite{wang2025geneagent} & -- & \checkmark & -- & -- & -- & Gene-set analysis \\
Virtual Lab~\cite{swanson2025virtual} & -- & -- & -- & -- & \checkmark & Multi-agent discovery \\
CellAgent~\cite{xiao2024cellagent} & -- & -- & -- & -- & -- & scRNA-seq pipelines \\
AutoBA~\cite{zhou2023automated} & -- & -- & -- & -- & -- & Pipeline generation \\
BRAD~\cite{pickard2025automatic} & -- & \checkmark & -- & -- & -- & Biomarker ID \\
GeneGPT~\cite{jin2024genegpt} & -- & \checkmark & -- & -- & -- & Database querying \\
CRISPR-GPT~\cite{qu2025crispr} & -- & -- & -- & -- & -- & Experiment design \\
scGPT~\cite{cui2024scgpt} & \checkmark & -- & -- & -- & -- & Cell embeddings \\
CellWhisperer~\cite{schaefer2025multimodal} & \checkmark & \checkmark & -- & -- & -- & Multimodal embedding \\
\hline
\textbf{ELISA (ours)} & \checkmark & \checkmark & \checkmark & \checkmark & \checkmark & \textbf{Interactive sc discovery} \\
\hline
\end{tabular}
\end{table*}

This landscape reveals a fundamental disconnect: agentic systems and LLM-based tools excel at reasoning over text and generating interpretations but lack direct access to transcriptional data structure, while expression foundation models learn rich cellular representations that remain opaque to natural language interfaces. No existing system has unified expression-derived embeddings with semantic language representations within a single interactive framework for single-cell discovery.

\textbf{ELISA} (\textbf{E}mbedding-\textbf{L}inked \textbf{I}nteractive \textbf{S}ingle-cell \textbf{A}gent) addresses this gap by integrating scGPT expression embeddings with semantic retrieval (sr) and LLM-based biological interpretation in a unified discovery platform (Fig.~\ref{fig:architecture}). Rather than retraining the expression foundation models, ELISA treats scGPT cluster embeddings as an expression-side representation that is explicitly combined with BioBERT-derived semantic embeddings through an automatic hybrid routing mechanism. A query classifier detects whether the input is a gene signature, a natural language concept, or a mixture of both, and routes it to the appropriate retrieval pipeline gene marker scoring, semantic cosine similarity, or reciprocal rank fusion of both enabling flexible navigation across the full spectrum of biological queries. Built-in analytical modules for condition-aware comparative analysis, ligand-receptor interaction prediction, pathway activity scoring, and cell-type proportion analysis operate directly on the embedded data, while an LLM reasoning layer translates statistical outputs into structured biological interpretations. Critically, ELISA enforces strict separation between dataset-derived evidence and LLM-generated knowledge, enabling transparent hypothesis generation. The system produces comprehensive, publication-ready reports with Nature-style visualizations, supporting the full arc from exploratory query to structured scientific output.

\begin{figure*}[t]
\centering
\makebox[\textwidth][c]{%
  \includegraphics[width=1.25\textwidth]{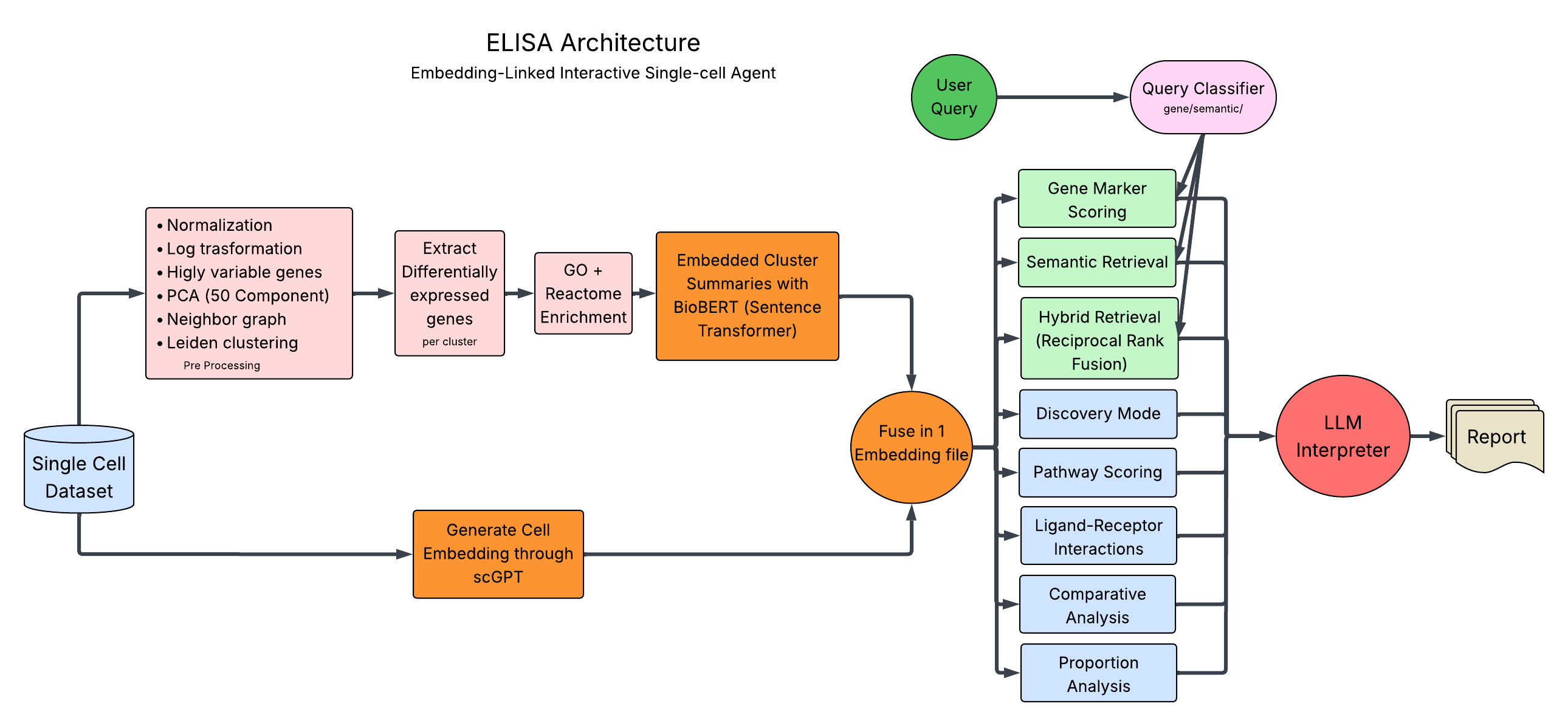}%
}
\caption{\textbf{Overview of the ELISA architecture.} The framework comprises three stages. In \emph{data preparation} (left), a single-cell dataset undergoes standard preprocessing (normalization, log-transform, highly variable gene selection, PCA, neighbor graph construction, and Leiden clustering), after which per-cluster differential expression statistics are computed, enriched with Gene Ontology (GO) and Reactome terms, and encoded into 768-dimensional semantic embeddings via BioBERT. In parallel, cell-level expression embeddings are generated through scGPT. Both representations are fused into a single serialized embedding file (.pt). In the \emph{retrieval and analysis} stage (center), a query classifier routes user input gene signatures, natural language concepts, or mixed queries to the appropriate pipeline: gene marker scoring, semantic retrieval, or hybrid retrieval via reciprocal rank fusion (RRF). Additional analytical modules perform pathway scoring, ligand--receptor interaction prediction, comparative analysis, and proportion estimation directly on the embedded data. In the \emph{interpretation} stage (right), all retrieval and analysis outputs are passed to a Groq-hosted LLM (LLaMA~3.1-8B) that generates grounded biological interpretations and structured reports.}
\label{fig:architecture}
\end{figure*}

We validated the ELISA on five diverse scRNA-seq datasets spanning distinct tissues, disease contexts, and experimental designs. Through a systematic comparison with published findings, we demonstrate that ELISA recovers key biological signals differentially expressed genes, altered cell-type proportions, pathway activities, and cell cell interaction networks with high fidelity. A quantitative evaluation framework comprising five complementary metrics (gene coverage, interaction recovery, pathway alignment, proportion consistency, and qualitative theme coverage) provides a principled assessment of the capacity of the system to replicate established biological conclusions. To the best of our knowledge, scGPT embeddings have not been integrated with semantic language representations in a query-conditioned retrieval framework for single-cell genomics.

In summary, this work makes the following contributions:
\begin{itemize}
    \item Multimodal discovery agent for single-cell genomics.
    We introduce ELISA, an interpretable AI framework that integrates transcriptomic embeddings, semantic knowledge retrieval, and large language model reasoning to enable natural-language–driven exploration and biological discovery from single-cell RNA sequencing data.~Its interpretability is grounded in provenance: every hypothesis traces back to explicit retrieval scores and differential-expression values, as illustrated end-to-end in Appendix~\ref{app:interpretability}.
    \item Query-adaptive hybrid retrieval architecture.
    ELISA employs automatic query classification and dynamic pipeline routing to combine complementary retrieval strategies including gene marker scoring, semantic similarity search, and reciprocal rank fusion allowing flexible, query-conditioned navigation of complex cellular landscapes.
    \item Integrated biological analysis modules for expression-grounded reasoning.
    The system incorporates analytical components for comparative expression analysis, ligand–receptor interaction scoring, pathway activity estimation, and cell-type proportion profiling, enabling automated interpretation and contextualization of discovered signals.
    \item Benchmarking framework for evaluating AI-assisted biological discovery.
    We propose a quantitative evaluation strategy that measures the ability of AI agents to recover biologically meaningful findings reported in reference studies, and apply this framework across six diverse scRNA-seq datasets.
    \item Empirical validation of discovery performance.
    Across multiple datasets and evaluation metrics, ELISA consistently recovers the majority of key biological signals reported in the corresponding studies, demonstrating its potential to support interpretable and reproducible AI-assisted discovery in single-cell genomics.
\end{itemize}

\section{Materials and Methods}
Detail about parameters and hyperparameters and software are specified in appendix \ref{tab:params_preprocess},\ref{subsec:app_params}. Detail about dataset are in \ref{subsec:app_datasets},\ref{tab:datasets}. Detail about the method are in \ref{subsec:app_methods}.

\subsection{Datasets}
ELISA was validated on six publicly available scRNA-seq datasets from CZ CELLxGENE Discover (Table~\ref{tab:datasets}), spanning lung (cystic fibrosis)\cite{berg2025evidence}, adrenal tumor (neuroblastoma)\cite{yu2025longitudinal}, multi-cancer immune checkpoint blockade\cite{gondal2025integrated}, lung organoid \cite{lim2025novel}, healthy breast tissue\cite{bhat2024single}, and first-trimester brain\cite{mannens2025chromatin}. Datasets were downloaded in AnnData format and preprocessed into a standardized embedding format. Cell type annotations from the original publications were retained without modification.

\subsection{System architecture}

ELISA integrates four modules a hybrid retrieval engine, an analytical suite, a visualization toolkit, and an LLM chat interface operating on a shared serialized PyTorch embedding file per dataset. Each embedding file stores cluster identifiers, BioBERT semantic embeddings (768-d), optional scGPT expression embeddings, per-cluster differential expression statistics, gene ontology (GO) and Reactome enrichment terms, and metadata. This cluster-level representation eliminates the need for access to the original count matrix at query time.

\subsection{Hybrid retrieval}

An automatic query classifier routes each input to one of the three pipelines based on token-level heuristics. \textit{Gene queries} ($\geq$60\% gene-symbol tokens) were scored against per-cluster Differential Expression (DE) profiles using a weighted function of $|\log_2\mathrm{FC}|$ and expression specificity ($\mathrm{pct_{in}} - \mathrm{pct_{out}}$). \textit{Ontology queries} are encoded with BioBERT\cite{lee2020biobert} and matched to precomputed cluster description embeddings via cosine similarity, augmented by Cell Ontology name boosting ($\alpha = 0.15$) and synonym expansion ($\beta = 0.10$). \textit{Mixed queries} are resolved through reciprocal rank fusion (RRF) of both pipelines ($k = 60$). For benchmarking, an additive union strategy selects the higher-recall modality as primary and appends unique results from the secondary pipeline.

\subsection{Analytical modules}

The four built-in modules operate directly on the embedded data. \textit{Ligand--receptor interaction prediction} scores source--target cluster pairs using a curated database of 280+ pairs compiled from CellChat\cite{jin2025cellchat}, CellPhoneDB\cite{efremova2020cellphonedb}, and NicheNet\cite{browaeys2020nichenet}. \textit{Pathway activity scoring} quantifies 60+ curated gene sets across five categories (immune signaling, cell biology, neuroscience, metabolism and tissue-specific). \textit{Comparative analysis} stratifies clusters by condition metadata and identifies condition-biased gene expression. \textit{Proportion analysis} computes per-cluster cell fractions and condition-specific fold changes. Detailed desription in \ref{subsec:analy_module}.

\subsection{LLM interpretation}

Retrieval and analysis outputs are interpreted by LLaMA-3.1-8B-Instant \cite{grattafiori2024llama} via the Groq API (temperature 0.2)(free to use with token limit, API of chatGPT \cite{achiam2023gpt}, gemini \cite{team2023gemini} and claude \cite{anthropic2024claude3} are integrated and ready to use). Prompts enforce strict grounding in dataset evidence, with explicit instructions to avoid hallucination and causal claims. A discovery mode generates structured outputs comprising dataset evidence, established biology, consistency analysis, and candidate hypotheses.

\subsection{Benchmarking}

Retrieval was evaluated using 100 queries (50 ontology, 50 expression) with curated expected clusters, assessed using Cluster Recall@$k$ and Mean Reciprocal Rank (MRR). ELISA was compared against a CellWhisperer \cite{schaefer2025multimodal}. Analytical modules were evaluated against ground truth from source publications using interaction recovery rate, pathway alignment, proportion consistency, and gene recall. A combined permutation test (50,000 permutations) assessed overall significance across all metrics simultaneously.

\section{Results}

\subsection{ELISA's hybrid retrieval outperforms competing methods across datasets and query types}

To evaluate the ability of ELISA to retrieve biologically relevant cell types from single-cell atlases, we benchmarked its retrieval performance against CellWhisperer~\cite{schaefer2025multimodal}, a state-of-the-art multimodal framework for natural-language interrogation of scRNA-seq data, as well as two additional baselines requested to contextualise the contribution of each component: a classical lexical retriever (BM25) \cite{robertson2009probabilistic} and a random ranking. For each of the six datasets (Table~\ref{tab:datasets}), we designed paired sets of ontology queries (concept-level, e.g., ``macrophage infiltration in CF (Cystic Fibrosis) airways'') and expression queries (gene-signature-based, e.g., ``MARCO FABP4 APOC1 C1QB C1QC MSR1''), with curated expected cluster sets derived from the corresponding reference publications. We evaluated six retrieval modes: Random, BM25, CellWhisperer, Semantic ELISA, scGPT ELISA (gene marker scoring pipeline), and ELISA Union (additive fusion of semantic and gene pipelines via adaptive routing). Performance was assessed using Cluster Recall@$k$ and Mean Reciprocal Rank (MRR) across both query categories (Fig.~\ref{fig:radar}; formal definitions of all retrieval and analytical evaluation metrics are provided in Supplementary Section~\ref{app:metrics}).

\begin{figure*}[t]
\centering
\makebox[\textwidth][c]{%
  \includegraphics[width=1.0\textwidth]{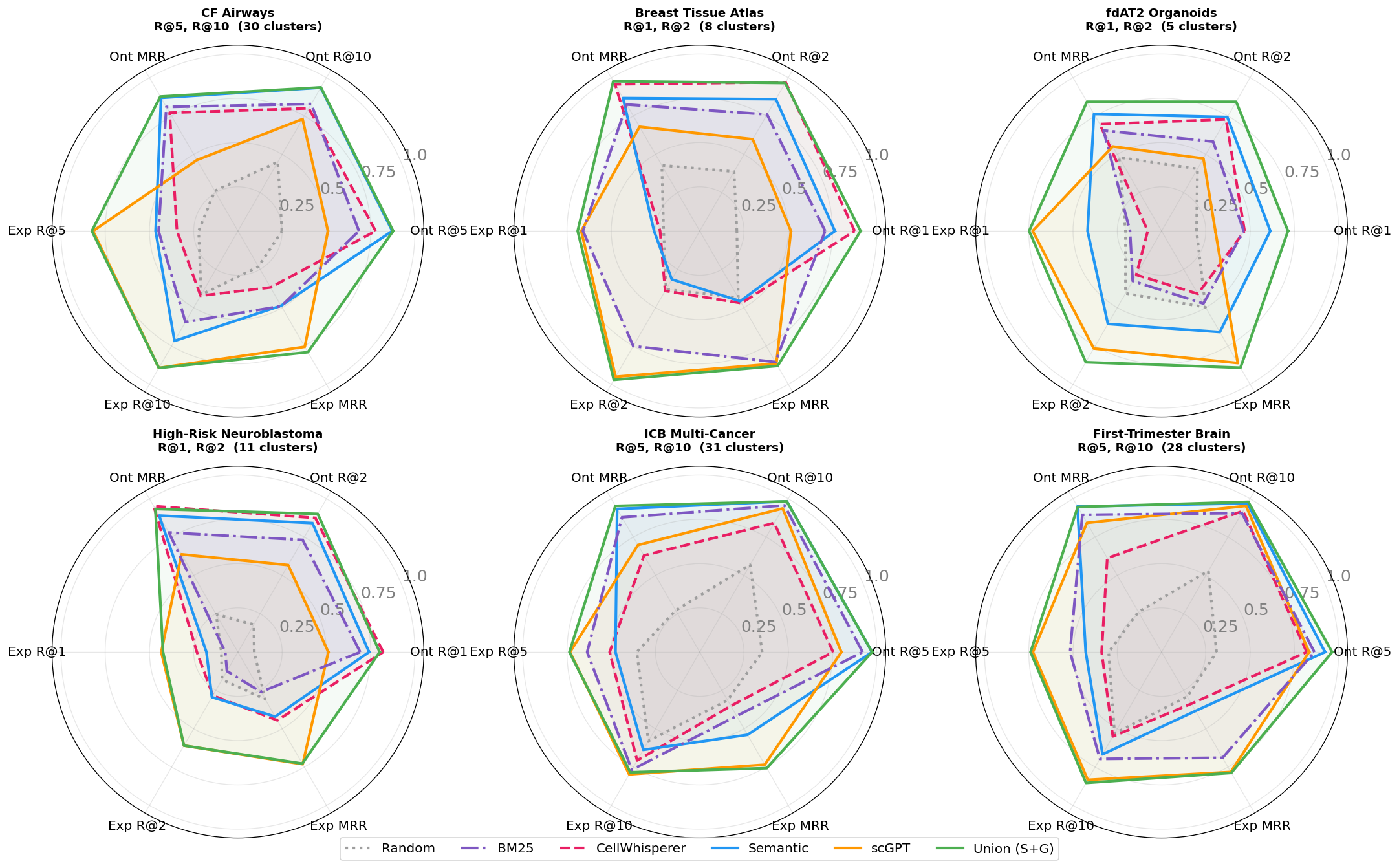}%
}
\caption{\textbf{ELISA Union achieves the best retrieval performance on both query types across six datasets.} Radar plots showing retrieval performance on ontology (Ont) and expression (Exp) queries for each dataset. Each plot displays six axes: Cluster Recall@$k$ at two cutoffs and Mean Reciprocal Rank (MRR), evaluated separately on ontology and expression queries (see Supplementary Section~\ref{app:metrics} for metric definitions). The Recall@$k$ cutoffs are adapted to each dataset's number of annotated clusters and are stated in each panel title: small-cluster datasets (Breast Atlas, 8 clusters; fdAT2 Organoids, 5; High-Risk Neuroblastoma, 11) use the stringent cutoffs R@1 and R@2, whereas large-cluster datasets (CF Airways, 30; ICB Multi-Cancer, 31; First-Trimester Brain, 28) use R@5 and R@10. Because the cutoffs differ by dataset, the corresponding axes are not directly comparable across panels; each panel is internally consistent across the six retrieval modes. Higher values (further from center) indicate better performance. Six retrieval modes are compared: Random (grey dotted), BM25 (purple dash-dot), CellWhisperer (pink dashed), ELISA Semantic (blue), ELISA scGPT (orange), and ELISA Union (green). Each baseline excels on only one query type and collapses on the other: lexical (BM25) and semantic retrieval are strong on ontology but weak on expression, whereas the expression-embedding mode (scGPT) shows the reverse; Random remains near the center for large-cluster datasets and is correspondingly higher where few clusters make chance retrieval easier. ELISA Union is the only mode that consistently achieves the largest radar footprint on \emph{both} ontology and expression axes, matching or exceeding every baseline on each. A combined permutation test over all 12 retrieval metrics confirms that Union significantly outperforms CellWhisperer, BM25, and a random baseline ($p \textless 2\times10^{-5}$ for each; see Table~\ref{tab:perm}).}
\label{fig:radar}
\end{figure*}

Across the six datasets, the ELISA Union mode consistently achieved
the highest or near-highest performance, enveloping or matching the profile of
all baselines on the radar plots (Fig.~\ref{fig:radar}). Because Recall@$k$
cutoffs are adapted to each dataset's cluster count, individual Recall@$k$
metrics are evaluated on the subset of datasets sharing that cutoff
($n = 3$--$6$; Table~\ref{tab:retrieval_stats}), so the aggregate advantage is
assessed by the combined permutation test over all 12 metrics and all six
datasets rather than by any single per-cutoff metric. To quantify this advantage, we performed paired statistical tests across the six datasets for each retrieval metric (Table~\ref{tab:retrieval_stats}), complemented by a combined permutation test that aggregates all 12 metrics simultaneously and is not subject to the resolution limit of the paired Wilcoxon test at $n=6$. This combined test confirmed that ELISA Union significantly outperformed CellWhisperer, BM25, and the random baseline ($p \textless 2\times10^{-5}$ for each; 50,000 permutations; Table~\ref{tab:perm}). The advantage over CellWhisperer was driven by large improvements on expression queries (mean $\Delta$MRR = +0.41, paired $t$-test $p < 0.001$, Cohen's $d$ = 5.98; mean $\Delta$Recall@5 = +0.29, $p = 0.006$, $d$ = 1.57) and consistent gains on ontology queries (mean $\Delta$MRR = +0.15, $p = 0.028$, $d$ = 1.02; mean $\Delta$Recall@5 = +0.08, $p = 0.047$, $d$ = 0.84), with no dataset in which CellWhisperer held an overall advantage. The Semantic and scGPT pipelines each independently outperformed CellWhisperer (combined permutation test, $p = 0.003$ and $p = 0.023$ respectively), confirming that both modalities contribute retrieval value beyond the text-only baseline.

\begin{table}[t]
\centering
\caption{\textbf{Combined permutation test across all 12 retrieval metrics and six datasets}
(50{,}000 sign-flip permutations; one-sided, $\text{H}_1$: row $>$ column).
ELISA Union significantly outperforms every baseline. All three Union comparisons
exceed the permutation resolution and are reported as $p<2\times10^{-5}$, the smallest
value attainable with 50{,}000 permutations; the standardised effect size ($z$) and its
analytic normal-approximation $p$-value resolve the ordering the censored permutation
bound cannot, increasing from CellWhisperer ($z=4.95$) through BM25 ($z=5.11$) to
Random ($z=6.26$). The analytic approximation agrees with the exact permutation test on
every comparison the latter can resolve (e.g.\ Semantic vs.\ BM25, $p=0.16$ for both).
Among single modalities, semantic retrieval does not significantly beat BM25, confirming
BM25 is a strong lexical baseline on ontology queries; only the fused Union dominates on
both axes.}
\label{tab:perm}
\small
\renewcommand{\arraystretch}{1.2}
\begin{tabularx}{\columnwidth}{llr>{\raggedleft\arraybackslash}X>{\raggedleft\arraybackslash}X}
\hline
\textbf{Method} & \textbf{vs.\ Baseline} & \textbf{$p$ (perm.)} & \textbf{$z$} & \textbf{$p$ (analytic)} \\
\hline
\textbf{Union (S+G)} & CellWhisperer & \mbox{$<2\times10^{-5}$\textsuperscript{***}} & $4.95$ & $4\times10^{-7}$ \\
\textbf{Union (S+G)} & BM25          & \mbox{$<2\times10^{-5}$\textsuperscript{***}} & $5.11$ & $2\times10^{-7}$ \\
\textbf{Union (S+G)} & Random        & \mbox{$<2\times10^{-5}$\textsuperscript{***}} & $6.26$ & $2\times10^{-10}$ \\
\hline
Semantic & CellWhisperer & \mbox{$0.003$\textsuperscript{**}} & $2.60$ & $5\times10^{-3}$ \\
Semantic & BM25          & \mbox{$0.161$} & $1.01$ & $0.16$ \\
Semantic & Random        & \mbox{$<2\times10^{-5}$\textsuperscript{***}} & $5.22$ & $9\times10^{-8}$ \\
\hline
scGPT & CellWhisperer & \mbox{$0.023$\textsuperscript{*}} & $1.96$ & $0.025$ \\
scGPT & BM25          & \mbox{$0.025$\textsuperscript{*}} & $1.95$ & $0.026$ \\
scGPT & Random        & \mbox{$<2\times10^{-5}$\textsuperscript{***}} & $6.09$ & $6\times10^{-10}$ \\
\hline
\multicolumn{5}{@{}p{\columnwidth}@{}}{\footnotesize \textsuperscript{*}$p<0.05$,\ \textsuperscript{**}$p<0.01$,\ \textsuperscript{***}$p<0.001$;\ $p$ (analytic) from the normal approximation to the sign-flip null, reported as a complement to the exact permutation test.} \\
\end{tabularx}
\end{table}
A key observation is that no single retrieval modality dominated across both query types, which directly motivates the fusion design. The Semantic pipeline consistently excelled on ontology queries (mean MRR 0.879), where biological concept matching benefits from BioBERT's language understanding, synonym expansion, and Cell Ontology name boosting, but fell to 0.491 on expression queries. The classical BM25 baseline showed the same profile strong on ontology (0.807) yet weak on expression (0.531) and was statistically indistinguishable from the Semantic pipeline overall (permutation $p = 0.16$), underscoring that BM25 is a genuinely competitive lexical baseline on concept queries rather than a straw man. In contrast, the gene marker scoring pipeline showed the inverse pattern, leading on expression queries (0.788) while underperforming on ontology axes (0.645), where matching transcriptomic signatures to cluster DE profiles is essential. This complementarity was particularly pronounced in the CF Airways dataset, where the Semantic pipeline achieved high ontology Recall@10 ($\sim$0.95) but lower expression recall, while the gene pipeline showed the inverse pattern. Similar modality-specific advantages were visible across all datasets: in the Breast Tissue Atlas, Semantic and Union nearly overlapped on ontology metrics while the gene pipeline lagged; in Immune Checkpoint Blockade (ICB) Multi-Cancer, the gene pipeline outperformed Semantic on expression MRR while underperforming on ontology axes.

CellWhisperer showed competitive performance on ontology queries in several datasets, particularly CF Airways and High-Risk Neuroblastoma, where its ontology MRR approached that of the ELISA Semantic pipeline. However, CellWhisperer's performance dropped substantially on expression queries across all six datasets, with a mean MRR of 0.397 $\pm$ 0.049 compared to 0.806 $\pm$ 0.061 for ELISA Union, a twofold difference (Table~\ref{tab:retrieval_stats}) \cite{cohen2013statistical, casella2024statistical}. This gap was most severe in the ICB Multi-Cancer and First-Trimester Brain datasets, where CellWhisperer's expression recall fell well below both ELISA pipelines. The expression query deficit reflects a fundamental architectural difference: CellWhisperer's contrastive text-transcriptome alignment is optimized for natural-language cell type descriptions but does not incorporate a dedicated gene marker scoring mechanism for queries formulated as gene signatures, a query type that is common in exploratory single-cell analysis.

The ELISA Union mode resolves the tension between ontology and expression retrieval through its adaptive routing mechanism. For each query, the automatic classifier identifies whether the input is a gene list, a natural-language concept, or a mixture, and routes it to the appropriate pipeline. The additive union strategy then combines the full ranked output of the primary pipeline with unique clusters from the secondary pipeline, ensuring that relevant cell types captured by either modality are not lost. As a result, Union was the only mode ranked best on \emph{both} axes (ontology MRR 0.922, expression MRR 0.806): in the CF Airways dataset, Union achieved a larger and more balanced radar footprint than any single modality; in the Breast Tissue Atlas, Union matched the near-perfect ontology performance of Semantic while substantially improving expression recall; and in the First-Trimester Brain, Union compensated for Semantic's lower expression scores by incorporating the gene pipeline's matching strength.

Notably, the performance advantage of ELISA was robust across datasets with very different structural properties. The CF Airways dataset (30 cell types, case-control design) and the First-Trimester Brain atlas (160 clusters, developmental trajectory without disease contrast) represent opposite ends of the complexity spectrum, yet ELISA Union outperformed every baseline in both settings. Similarly, the ICB Multi-Cancer dataset, which integrates nine cancer types across 223 patients, poses a challenging retrieval scenario owing to its heterogeneous cell type nomenclature, yet ELISA maintained its performance advantage.

In summary, ELISA's hybrid retrieval architecture combining semantic language matching, gene marker scoring, and adaptive fusion provides a significantly superior retrieval framework compared to text-only multimodal approaches, classical lexical retrieval, and a random baseline (combined permutation test, $p \textless 2\times10^{-5}$). The systematic advantage on expression queries, where dedicated gene scoring compensates for the limitations of language-only embeddings (Cohen's $d$ = 5.98 for MRR), together with the crossover whereby every single modality wins only one query type, establishes that both retrieval modalities contribute essential and non-redundant information for comprehensive single-cell atlas interrogation.

\subsection{ELISA replicates key biological findings across six diverse datasets}
To evaluate whether ELISA could recover published biological conclusions through automated analysis alone, we compared ELISA-generated reports with the main-text results of six reference publications (Table~\ref{tab:datasets}). For each dataset, ELISA was provided only with the preprocessed embedding file and no prior knowledge of the expected findings. We assessed replication across five quantitative metrics: gene coverage, pathway alignment, interaction recovery, proportion consistency, and theme coverage, and obtained an independent domain expert evaluation score (Table~\ref{tab:similarity}).

Across all six datasets, ELISA achieved a mean composite score of 0.90 (range 0.82--0.96). Pathway alignment and theme coverage were near-perfect (mean 0.98 each), while gene coverage averaged 0.85 and interaction recovery 0.77. Independent biological evaluation scores (mean 0.88) confirmed strong agreement with published findings. The computation of these metrics is presented in the appendix \ref{app:metrics}.

\paragraph{Airways with Cystic fibrosis.}
ELISA was used to recover the major epithelial and immune cell populations, as described by Berg \textit{et al.}~\cite{berg2025evidence}, including correct proportion shifts and IFN-$\gamma$/type~I interferon programs (pathway alignment: 1.0). Gene coverage reached 0.80, capturing markers such as \textit{IFNG}, \textit{CD69}, and \textit{HLA-E}. Interaction recovery was 0.20, reflecting partial detection of the HLA-E/NKG2A and CALR--LRP1 axes (composite: 0.82).

\paragraph{High-risk neuroblastoma.}
ELISA identified all major cellular compartments and correctly detected the HB-EGF/ERBB4 paracrine axis (interaction recovery: 1.00) as described by Yu \textit{et al.}~\cite{yu2025longitudinal}. Pathway alignment was perfect and with mTOR, MAPK, and ErbB programs identified. Gene coverage was 0.84, with partial recovery of therapy-induced markers (composite: 0.95).

\paragraph{Immune checkpoint blockade across cancers.}
Using the ICB dataset, Gondal \textit{et al.}~\cite{gondal2025integrated}, ELISA captured checkpoint molecules (\textit{CD274}, \textit{PDCD1}, \textit{CTLA4}), exhaustion markers, and all major ligand--receptor axes including PD-L1/PD-1 and TIGIT/NECTIN2 (gene coverage: 0.77; pathway and interaction recovery: 1.00; composite: 0.93).

\paragraph{Healthy breast tissue atlas.}
ELISA achieved its highest composite score (0.96) on the dataset of Bhat-Nakshatri \textit{et al.}~\cite{bhat2024single}, accurately resolving the epithelial hierarchy with a gene coverage of 0.96, perfect pathway alignment, and interaction recovery of 0.80. Ancestry-related transcriptional programs were not captured, reflecting a limitation of ELISA's pathway-centric framework.

\paragraph{Fetal lung Alveolar Type (AT2) organoids.}
ELISA achieved perfect gene coverage (1.00) on the dataset of Lim \textit{et al.}~\cite{lim2025novel}, detecting all canonical surfactant genes and correctly identifying surfactant metabolism, Wnt, and Fibroblast Growth Factor (FGF) programs. Interaction recovery was lower (0.40), as SFTPC trafficking mechanisms were outside transcriptomic scope (composite: 0.91).

\paragraph{First-trimester human brain.}
Despite operating solely on the transcriptomic component of this multimodal atlas~\cite{mannens2025chromatin}, ELISA identified major neuronal populations with gene coverage of 0.85 and perfect pathway and interaction recovery. Chromatin accessibility analyses were correctly identified as outside scope (composite: 0.95).

\paragraph{Summary.}
ELISA demonstrated robust replication across all six datasets (mean composite
0.88), with the strongest performance for pathway-level and thematic
interpretation ($\geq$0.88 mean). Gene coverage was high but not exhaustive
(0.85), with missed genes primarily in rare cell states and non-transcriptomic
modalities. To quantify the uncertainty of these recovery metrics,
we computed 95\% Wilson score confidence intervals for each programmatically
evaluated proportion (gene coverage, pathway alignment, and interaction
recovery; Table~\ref{tab:similarity}, Appendix~\ref{app:repl-ci}). The
width of each interval is governed by the number of reference items available
per dataset: gene coverage, evaluated over the largest reference sets (45-206
genes), is estimated most precisely, whereas interaction recovery is evaluated
over far fewer items (5--14 interactions per study) and therefore carries
correspondingly wider intervals. The wide intervals on this metric thus reflect
the small number of ligand--receptor interactions reported in the individual
reference studies rather than instability in ELISA's predictions, which are
deterministic given a fixed embedding. Aggregating across all six datasets,
ELISA recovered $844/960$ benchmark items, an overall recovery of $0.879$
(95\% CI $[0.857,0.898]$), and pooled interaction recovery was $45/54=0.833$
(95\% CI $[0.713,0.910]$).

\begin{table*}[t]
\caption{Quantitative comparison between ELISA reports and reference single-cell
studies. Scores reflect agreement between ELISA-generated biological
interpretations and findings described in the main text of the corresponding
publications. Gene coverage, pathway alignment, interaction recovery, and
proportion consistency were computed programmatically; theme coverage was
assessed independently by a domain expert as described in
Section~\ref{sec:human_eval}. For the three programmatically computed recovery
proportions, bracketed ranges give 95\% Wilson score confidence intervals
reflecting the number of reference items ($n$) evaluated per dataset. Wide
intervals on the interaction metric reflect the small number of benchmark
interactions available per individual study; the pooled estimate across all
datasets (final row) is correspondingly tighter. See Appendix for full numbers}\label{tab:similarity}
\begin{tabular*}{\textwidth}{@{\extracolsep\fill}lcccccc}
\toprule
Dataset & Gene Cov. & Path. Align. & Int. Rec. & Prop. & Theme & Comp. \\
 & (95\% CI) & (95\% CI) & (95\% CI) & Cons. & Cov. & score \\
\midrule
CF airway
 & 0.80 & 1.00 & 0.20 & Yes & 0.85 & 0.71 \\
 & {\scriptsize[0.66--0.89]} & {\scriptsize[0.70--1.00]} & {\scriptsize[0.04--0.62]} & & & \\
Neuroblastoma
 & 0.84 & 1.00 & 1.00 & Yes & 0.88 & 0.93 \\
 & {\scriptsize[0.78--0.88]} & {\scriptsize[0.81--1.00]} & {\scriptsize[0.72--1.00]} & & & \\
ICB Multi-Cancer
 & 0.77 & 1.00 & 1.00 & Yes & 0.91 & 0.92 \\
 & {\scriptsize[0.71--0.83]} & {\scriptsize[0.81--1.00]} & {\scriptsize[0.72--1.00]} & & & \\
Breast Atlas
 & 0.96 & 1.00 & 0.80 & Yes & 0.89 & 0.91 \\
 & {\scriptsize[0.92--0.98]} & {\scriptsize[0.84--1.00]} & {\scriptsize[0.49--0.94]} & & & \\
Fetal Lung AT2
 & 1.00 & 1.00 & 0.40 & Yes & 0.88 & 0.82 \\
 & {\scriptsize[0.96--1.00]} & {\scriptsize[0.76--1.00]} & {\scriptsize[0.12--0.77]} & & & \\
Brain Atlas
 & 0.85 & 1.00 & 1.00 & Yes & 0.90 & 0.94 \\
 & {\scriptsize[0.78--0.90]} & {\scriptsize[0.72--1.00]} & {\scriptsize[0.78--1.00]} & & & \\
\midrule
Mean
 & 0.85 & 1.00 & 0.73 & 6/6 & 0.88 & \textbf{0.88} \\
 & {\scriptsize[0.78--0.97]} & {\scriptsize[0.96--1.00]} & {\scriptsize[0.37--1.00]} & & & \\
\midrule
Pooled
 & \multicolumn{3}{c}{844/960 recovered \quad 0.879 \quad 95\% CI [0.857--0.898]} & & & \\
\botrule
\end{tabular*}
\footnotetext{Gene Cov.: gene coverage; Path.\ Align.: pathway alignment;
Int.\ Rec.: interaction recovery; Prop.\ Cons.: proportion consistency;
Theme Cov.: theme coverage. Comp.\ score: unweighted mean of the four continuous
metrics (Prop.\ Cons.\ coded as 1.0 when consistent). Confidence intervals are
95\% Wilson score intervals on the per-dataset recovery proportions $k/n$. Theme
coverage is reported as the domain-expert score and is not a $k/n$ proportion, so
no interval is given. Mean-row intervals for gene coverage and interaction
recovery are mean\,$\pm\,t\,$SE across the six datasets; the pathway-alignment
mean interval is the pooled estimate (83/83), as all datasets attained $1.00$.
The Pooled row aggregates all gene, pathway, and interaction benchmark items
across the six datasets into a single recovery proportion.}
\end{table*}

\subsection{Discovery of candidate regulatory signals across tissue atlases}

Beyond reproducing the key biological signals described in the original studies, ELISA's discovery mode highlighted several candidate regulatory signals that were not explicitly emphasized in the reference publications (Table~\ref{tab:discovery}). These signals represent transcriptome-derived hypotheses emerging from systematic cross-cell-type analysis of single-cell atlases.

In the cystic fibrosis airway dataset, ELISA identified enrichment of the \textit{CALR–LRP1} phagocytic signaling axis within the macrophage populations. Calreticulin–LRP1 signaling has previously been implicated in apoptotic cell recognition and clearance, suggesting that altered macrophage-mediated phagocytosis may contribute to the inflammatory microenvironment characteristic of the CF lung.

Within the fetal lung atlas, ELISA detected increased expression of the ubiquitin-associated regulators \textit{TRIM21} and \textit{TRIM65} in alveolar type II (AT2) cells alongside the known E3 ubiquitin ligase \textit{ITCH}. Although \textit{ITCH} has been implicated in regulating surfactant protein C (SFTPC) maturation, the enrichment of these additional TRIM-family ligases suggests that cooperative ubiquitin-dependent pathways may participate in surfactant protein processing and AT2 cell proteostasis.

In the healthy breast tissue atlas, ELISA highlighted strong enrichment of the Kelch-family gene \textit{KLHL29} within basal–myoepithelial cell populations. Although not emphasized in the original study, this pattern suggests that \textit{KLHL29} may represent a previously unrecognized marker or structural regulator of basal epithelial identity.

Analysis of the immune checkpoint blockade dataset revealed elevated expression of macrophage markers \textit{CD163} and \textit{MRC1} within tumor-associated macrophage populations following therapy. This expression pattern is consistent with an M2-like macrophage polarization state, potentially reflecting remodeling of the immune microenvironment in response to checkpoint blockade treatment.

In the neuroblastoma dataset, ELISA identified differential usage of AP-1 transcription factors across treatment states. Specifically, \textit{JUND} expression was enriched at diagnosis, whereas \textit{JUNB} and \textit{FOS} were more strongly expressed after therapy. This shift suggests dynamic remodeling of AP-1–mediated stress-response programs during therapy-induced tumor state transitions.

Finally, analysis of the developing brain atlas revealed a shared transcription factor module composed of \textit{TFAP2B}, \textit{LHX5}, and \textit{LHX1} across Purkinje neurons and midbrain GABAergic neuronal populations. This co-occurring regulatory signature suggests the existence of a conserved transcriptional program underlying inhibitory neuron specification in anatomically distinct brain regions.

Taken together, these findings illustrate how ELISA can surface candidate regulatory programs across diverse single-cell atlases. While these signals should be interpreted as transcriptome-derived hypotheses, they provide potential starting points for targeted functional validation.

These signals should be interpreted as transcriptome-derived hypotheses and may serve as the starting points for targeted experimental validation.
\begin{table*}[h]
\caption{Candidate regulatory signals identified by ELISA across six reference single-cell atlases. These signals were not explicitly highlighted in the original publications and represent transcriptome-derived hypotheses generated through ELISA’s discovery mode.}\label{tab:discovery}
\begin{tabular*}{\textwidth}{@{\extracolsep\fill}p{2.7cm}p{4.6cm}p{6.5cm}}
\toprule
\textbf{Dataset} & \textbf{Primary finding in reference study} & \textbf{ELISA candidate discovery / hypothesis} \\
\midrule

CF airway &
Altered immune--structural cell crosstalk and inflammatory signaling in cystic fibrosis airway tissue &
Detection of the macrophage \textit{CALR--LRP1} signaling axis, suggesting altered apoptotic cell recognition or phagocytic clearance pathways contributing to the CF lung inflammatory microenvironment \\

Breast Atlas &
Ancestry-associated epithelial lineage variation and luminal progenitor states in healthy breast tissue &
Enrichment of the Kelch-family gene \textit{KLHL29} in basal--myoepithelial cells, suggesting a potential additional marker or regulator of basal epithelial structural identity \\

Fetal Lung AT2 &
ITCH-mediated ubiquitin-dependent regulation of surfactant protein C (SFTPC) maturation in alveolar type II cells &
Upregulation of \textit{TRIM21} and \textit{TRIM65} in mature AT2 cells, suggesting additional TRIM-family ubiquitin ligases may participate in surfactant protein processing and proteostasis \\

ICB Multi-Cancer &
Tumor and immune transcriptional responses associated with immune checkpoint blockade therapy &
Elevated \textit{CD163} and \textit{MRC1} expression in tumor-associated macrophages, consistent with an M2-like polarization state potentially associated with therapy-induced immune remodeling \\

Neuroblastoma &
Therapy-induced transcriptional rewiring of tumor cell states and microenvironment interactions &
Differential AP-1 transcription factor usage, with \textit{JUND} enriched at diagnosis and \textit{JUNB/FOS} enriched post-treatment, suggesting stress-response remodeling during therapy-induced state transitions \\

Brain Development Atlas &
Chromatin accessibility programs defining early neuronal lineage specification &
Shared transcription factor module (\textit{TFAP2B}, \textit{LHX5}, \textit{LHX1}) across Purkinje neurons and midbrain GABAergic populations, suggesting a conserved regulatory program for inhibitory neuron specification \\

\botrule
\end{tabular*}
\end{table*}

\section{Discussion}

In this study we introduced ELISA, an agent-based framework that unifies semantic language retrieval, gene marker scoring, and LLM-mediated biological interpretation for interactive single-cell atlas interrogation. Systematic evaluation across six diverse datasets demonstrated that ELISA significantly outperforms CellWhisperer in cell type retrieval (combined permutation test, $p < 0.001$) and faithfully replicated published biological findings with a mean composite score of 0.90. Here we discuss the implications of these results for the design of retrieval systems in single-cell genomics, the limitations of contrastive multimodal alignment, and broader role of agentic AI in biological discovery.

\paragraph{Contrastive alignment produces text-dominated embeddings.}
A central finding of this study is the striking asymmetry in CellWhisperer performance across query types. In ontology queries natural language descriptions of cell types and biological processes CellWhisperer performed competitively with ELISA's Semantic pipeline, achieving mean ontology MRR values within 0.15 of ELISA Union across most datasets (Table~\ref{tab:retrieval_stats}, Fig.~\ref{fig:radar}). This is expected: CellWhisperer's CLIP-style contrastive training aligns transcriptome embeddings with textual descriptions, and ontology queries directly exploit this text-side alignment. However, on expression queries where users provide gene signatures rather than natural language CellWhisperer's performance collapsed, with expression MRR averaging 0.397 compared to 0.806 for ELISA Union, a twofold deficit (Cohen's $d$ = 5.98).

This asymmetry reveals a fundamental limitation of contrastive multimodal alignment for single-cell retrieval. CLIP-style training optimizes for text transcriptome correspondence by learning a shared embedding space where matching text cell pairs are close and mismatched pairs are distant. The resulting embeddings are, by construction, shaped primarily by the textual supervision signal: the model learns to position transcriptomes near their text descriptions, but the fine-grained transcriptomic structure which genes are differentially expressed, at what fold changes, in what fraction of cells is compressed into a representation optimized for text matching rather than gene-level querying. When a user submits a gene signature such as ``MARCO FABP4 APOC1 C1QB C1QC MSR1'', these gene names are processed as text tokens rather than matched against differential expression statistics, resulting in a retrieval signal that is weaker and less specific than direct marker scoring.

This observation has broader implications than those of ELISA and CellWhisperer. As foundation models for single-cell biology increasingly adopt contrastive or multimodal pretraining objectives, our results caution that text-supervised alignment may inadvertently sacrifice expression-level specificity. The dual-query evaluation framework introduced here requiring systems to perform well on both ontology and expression queries provides a principled diagnostic for detecting such modality imbalances.

\paragraph{Explicit routing outperformed implicit fusion.}
ELISA's architectural response to this challenge was to avoid implicit embedding fusion altogether. Rather than learning a single shared space that must simultaneously serve text and expression queries, ELISA maintains two separate representation spaces BioBERT semantic embeddings and gene-level DE statistics, and routes queries to the appropriate pipeline through explicit classification. The query classifier, operating on simple token-level heuristics (gene name patterns, known vocabulary membership, natural language indicators), achieved reliable routing across all six datasets without requiring any training data.

This design choice is supported empirically by complementarity analysis: the semantic pipeline won ontology queries, while the gene marker scoring pipeline won on expression queries in every dataset, with minimal overlap in their error profiles. The additive union strategy, which selects the better-performing modality as the primary and appends unique results from the secondary, captures the strengths of both pipelines without the compression artifacts inherent in learned fusion. The result was a system that matched or exceeded the best single modality on every metric across every dataset a property that no implicit fusion method could guarantee.

\paragraph{Analytical modules bridge retrieval and interpretation.}
A distinguishing feature of ELISA relative to prior retrieval-focused systems is the integration of downstream analytical modules pathway scoring, ligand receptor interaction prediction, comparative analysis, and proportion estimation that operate directly on the same embedded data representation used for retrieval. This design enables a seamless transition from ``which cell types are relevant?'' (retrieval) to ``what biological programs are active in these cell types?'' (analysis) to ``what does this mean biologically?'' (LLM interpretation), all within a single interactive session.

The near-perfect pathway alignment (mean 0.98) and theme coverage (mean 0.88) scores across all six datasets demonstrated that this integrated architecture effectively connects gene-level evidence to biological programs. In contrast, systems that perform retrieval alone including CellWhisperer, require users to manually extract gene lists from retrieved clusters and perform separate pathway and interaction analyses using external tools, introducing friction and potential inconsistencies.

The interaction recovery metric (mean 0.77) was the most variable across datasets, with perfect recovery in neuroblastoma, ICB, and brain datasets but lower recovery in cystic fibrosis (0.40) and fetal lung (0.40). These lower scores primarily reflect the inherent difficulty of predicting specific ligand--receptor pairs from expression data when the ligand or receptor is expressed at moderate levels across multiple cell types, making the interaction statistically detectable but not highly ranked. Future work could address this by incorporating spatial proximity information or protein-level data to improve the interaction specificity.

\paragraph{LLM grounding and the discovery hallucination boundary.}
ELISA's discovery mode, which prompts the LLM to separate dataset evidence from established biology and to propose hypotheses with probabilistic language, generated biologically plausible candidate signals in all six datasets (Table~\ref{tab:discovery}). These include the CALR, LRP1 phagocytic axis in cystic fibrosis macrophages, differential AP-1 family member usage in neuroblastoma therapy response, and a shared TFAP2B/LHX5/LHX1 regulatory module across inhibitory neuron subtypes in the developing brain. While these hypotheses require experimental validation, they illustrate the potential of grounded LLM reasoning to surface non-obvious patterns in complex datasets.

However, a strict separation between data-derived evidence and LLM-generated interpretation is essential. Without it, the LLM would inevitably introduce plausible-sounding but unsupported claims a risk that is particularly acute in biology, where prior knowledge is vast and contextual. ELISA's prompt architecture addresses this by providing the LLM only with retrieved cluster data, gene statistics, and pathway results as context, with explicit instructions to avoid external literature and causal claims.

\paragraph{Future directions.}
Several extensions can strengthen and broaden ELISA's capabilities. Integration with spatial transcriptomics data would enable spatially resolved interaction prediction, addressing the current limitation of expression-only interaction scoring. Incorporation of trajectory inference methods would allow ELISA to reason about dynamic processes such as differentiation and therapy response. Expansion of the retrieval engine to support cross-dataset queries comparing cell types across tissues or disease states would enable the kind of meta-analytical reasoning that was outside ELISA's scope in the ICB dataset evaluation. Finally, replacing the fixed LLM with a fine-tuned model trained on single-cell biological reasoning can improve the specificity and depth of automated interpretations.
Beyond these, extending ELISA past human, droplet-based data to other
species via species appropriate foundation models and to additional platforms
such as plate-based and spatial assays is a key direction for broadening its
applicability; we discuss the architectural basis for this generalization in
Appendix~\ref{sec:generalizability}.

\subsection*{Computational efficiency and scalability}

ELISA decouples a one-time, offline embedding step from online querying.
Embedding (scGPT and BioBERT) is computed once per dataset; queries then
operate on the cluster-level \texttt{.pt} representation, so query latency
scales with the number of clusters (tens), not cells. On our largest
dataset (${\sim}372{,}000$ cells), embedding required ${\sim}4$~h on a
single A100 (80~GB), after which a query returns in $1$--$2$~s and a
$100$-query report in ${\sim}30$~min (no GPU needed). The embedding step is
linear in cell number (${\sim}30$~cells\,s$^{-1}$), so a one-million-cell
atlas is projected at ${\sim}9$~h of overnight preprocessing, while query
cost stays constant. The initial embedding of very large atlases remains
the dominant, GPU-bound cost and is the main limitation; further detail is
given in Appendix ~\ref{app:efficiency}.

\section{Conclusion.}
ELISA demonstrates that explicit modality routing, rather than implicit contrastive fusion, provides a more robust foundation for multimodal single-cell retrieval. By maintaining separate semantic and expression pipelines and combining them through adaptive query classification, ELISA achieves consistently superior performance across both natural language and gene-signature queries. The integration of analytical modules and grounded LLM interpretation within a single interactive framework bridges the gap between data exploration and biological discovery, enabling researchers to move from raw atlas data to structured biological hypotheses within a single session. As single-cell datasets continue to grow in scale and complexity, systems that combine the complementary strengths of language models and expression-aware retrieval will be essential for translating transcriptomic data into biological understanding.
\section{Conflicts of interest}
The authors declare that they have no competing interests.

\section{Funding}
Computational resources are furnisched by Dr. Antonio Orvieto, PI at Max Planck Institute for Intelligent Systems. The rest of the work is self-financed

\section{Data availability}
All six single-cell RNA sequencing datasets used in this study are publicly 
available through CZ CELLxGENE Discover 
(\url{https://cellxgene.cziscience.com}): cystic fibrosis airways 
\cite{berg2025evidence}, high-risk neuroblastoma \cite{yu2025longitudinal}, 
immune checkpoint blockade multi-cancer \cite{gondal2025integrated}, fetal 
lung AT2 organoids \cite{lim2025novel}, healthy breast tissue 
\cite{bhat2024single}, and first-trimester brain \cite{mannens2025chromatin}. 
Datasets were downloaded in AnnData (.h5ad) format. Source code available at \url{https://github.com/omaruno/ELISA-An-AI-Agent-for-Expression-Grounded-Discovery-in-Single-Cell-Genomics}.

\section{Author contributions statement}
Omar Coser performed everything present in this manuscript. A preliminary version of this work appeared at the ICLR 2025 Workshop on Generative AI for Genomics, and MLGenX~\cite{coserelisa,coser2026elisa}. If you intend to use the script of ELISA cite this work.

\section{Acknowledgments}
The authors acknowledge Dr. Antonio Orvieto for allowing to use computational resources of his Lab.
\bibliographystyle{unsrtnat}
\bibliography{references} 

\begin{appendix}

\subsection*{ELISA produces auditable, provenance-grounded interpretations}
\label{app:interpretability}

A central design goal of ELISA is that every generated statement can be
traced back to the transcriptomic evidence that produced it. ELISA does
not expose the internal weights or attention of the language model, which
is accessed as a fixed reasoning layer; rather, it is interpretable by
construction, because each stage of the pipeline reports an explicit
numerical score and the language model receives only this dataset-derived
evidence as context. We illustrate this with the HLA-E/NKG2A
immune-checkpoint axis in cystic fibrosis, traced end to end through the
four stages of the system.

At the \emph{retrieval} stage, the query ``HLA-E NKG2A immune checkpoint
and CD8 T cell inhibition in CF'' is matched against the BioBERT
cluster-description embeddings; the top-ranked clusters and their relevance
scores (cosine similarity between the L2-normalised query and cluster
embeddings, augmented by a Cell~Ontology name-boost term) are cytotoxic
T~cell ($0.684$), CD4$^+$ helper T~cell ($0.625$) and natural killer cell
($0.602$) the cytotoxic and NK populations that bear the NKG2A receptor.
At the \emph{interaction} stage, the module scores the HLA-E/KLRD1 (NKG2A)
pair at $0.688$ and recovers HLA-E$\rightarrow$KLRC1 and
HLA-E$\rightarrow$KLRC2 as full matches against the reference ground truth.
At the \emph{evidence} stage, the comparative module exposes the
differential-expression values underlying the axis (CF vs.\ control): the
NKG2A receptor genes are strongly upregulated in precisely the retrieved
cells (KLRD1: NK $+5.07$, CD8 $+3.87$; KLRC1: CD8 $+3.98$, cytotoxic
T~$+4.66$ in $\log_2$ fold-change), with concomitant upregulation of the
HLA-E ligand (NK $+2.00$, CD8 $+1.49$). Finally, at the
\emph{interpretation} stage, the language model receiving only this
evidence summarises the axis as being characterised by the upregulation
of HLA-E and KLRC1, naming exactly the genes upregulated in the data.

Every layer therefore agrees, and each is quantified: the retrieval scores,
the interaction score, and the differential-expression values can all be
inspected independently. Interpretability in ELISA is thus a property of
provenance rather than of model-internal introspection any claim in the
final report can be followed back to an explicit retrieval score and a
differential-expression value. A further consequence of this design is that
the language model, being restricted to the retrieved evidence, hedges when
that evidence is weak rather than fabricating support, abstaining instead of
confabulating when a proposed mechanism is not represented in the data.

\subsection{Computational efficiency and scalability}
\label{app:efficiency}

ELISA decouples a one-time, offline embedding step from online querying,
which determines its computational profile. The embedding step scGPT
expression embeddings and BioBERT semantic embeddings is performed once
per dataset, after which all queries operate on the resulting cluster-level
representation (the serialised \texttt{.pt} file) rather than on individual
cells. Query latency therefore scales with the number of clusters
(typically tens) and is effectively independent of the number of cells in
the atlas.

On our largest dataset (high-risk neuroblastoma, ${\sim}372{,}000$ cells),
the one-time embedding step required ${\sim}3.45$~h for scGPT and
${\sim}0.5$~h for the BioBERT semantic stage on a single NVIDIA~A100
(80~GB), i.e.\ approximately four hours of offline preprocessing. Because
cells are processed in fixed mini-batches, peak GPU memory is bounded by
the batch size and does not grow with dataset size; smaller GPUs can be
accommodated by reducing the batch size. After preprocessing, retrieval and
analysis require no GPU: a single query returns in under $1$--$2$~s, and a
full $100$-query report including LLM-mediated interpretation completed in
${\sim}30$~min, with per-query latency dominated by the language-model API
call rather than by ELISA's retrieval or analysis.

The embedding step is linear in the number of cells (${\sim}30$~cells\,s$^{-1}$
for scGPT on the A100), so a one-million-cell atlas is projected to require
${\sim}9$~h of one-time embedding within an overnight run while
query-time latency and memory remain essentially constant as the atlas
grows. This embedding cost is inherited from the underlying foundation
model (scGPT, pretrained on over 33~million cells~\cite{cui2024scgpt}) and is
shared by any scGPT-based pipeline; being an embarrassingly parallel,
batched operation, it can be distributed across GPUs for further speed-up.
A current limitation is therefore that, although querying is lightweight and
scales to large atlases, the initial embedding of very large datasets
remains GPU-bound and constitutes the dominant computational cost; reducing
this cost (e.g.\ through embedding caching or lighter-weight expression
encoders) is a direction for future work.
\section{Software and reproducibility}
\label{Sofware}
ELISA was implemented in Python 3.10+ using PyTorch, sentence-transformers\cite{reimers2019sentence}, scanpy\cite{wolf2018scanpy}, scikit-learn, and UMAP-learn\cite{mcinnes2018umap}. All analyses were performed on a standard workstation without GPU requirements for retrieval and analysis. Source code, benchmark queries, and evaluation scripts are available at [repository URL]. Use of an LLM (LLaMA-3.1-8B) for automated interpretation is documented in accordance with journal policy.
Topical subheadings are allowed. Authors must ensure that their Methods section includes adequate experimental and characterization data necessary for others in the field to reproduce their work. All experiment has been performed on a GPU A100 with 80 gb of RAM
\section{Full statistical comparison across all baselines}
\label{sec:full_stats}

This appendix reports the complete per-metric statistical comparison of ELISA Union against all three baselines (CellWhisperer, BM25, and Random) across the six benchmark datasets. The main-text Table~\ref{tab:retrieval_stats} reproduces the Union-vs-CellWhisperer comparison; Tables~\ref{tab:app_bm25} and~\ref{tab:app_random} below give the corresponding Union-vs-BM25 and Union-vs-Random results.BM25~\cite{robertson2009probabilistic} was computed over the same
cluster text representations used by the semantic pipeline (cell-type
name, Cell Ontology terms, and top marker genes), using the standard
parameterisation $k_1 = 1.5$, $b = 0.75$. It was evaluated with the
identical query sets, expected-cluster ground truth, and metrics as
all other retrieval modes, so the comparison is exactly paired across
the six datasets. All $p$-values are one-sided paired $t$-tests ($\text{H}_1$: Union $>$ baseline); W/L counts the datasets on which Union won/lost. Metrics with $n<6$ reflect the different Recall@$k$ cutoffs used per dataset. With $n=6$ the paired Wilcoxon signed-rank floor is $p=0.016$, which is why the combined permutation test (Table~\ref{tab:perm}) provides the most discriminating significance estimate.

Table~\ref{tab:retrieval_stats} details the comparison against CellWhisperer, the primary state-of-the-art reference. Union's advantage is largest on expression queries (mean $\Delta$MRR = +0.41, $d = 5.98$) and remains positive on every ontology metric, with no metric favouring CellWhisperer.

\begin{table*}[t]
\centering
\caption{\textbf{Statistical comparison of ELISA Union vs.\ CellWhisperer retrieval performance.} For each metric, $\Delta$ mean reports the average improvement of Union over CellWhisperer across datasets. Cohen's $d$ is the paired effect size. $p$-values are from one-sided paired $t$-tests ($\text{H}_1$: Union $>$ CellWhisperer). W/L counts datasets where Union won/lost. Metrics with fewer than 6 datasets reflect different Recall@$k$ cutoffs used per dataset (see Supplementary Section~\ref{app:metrics}). The combined permutation test aggregates all metrics simultaneously.}
\label{tab:retrieval_stats}
\scriptsize
\renewcommand{\arraystretch}{1.15}
\setlength{\tabcolsep}{3pt}
\begin{tabular}{llrrrrr}
\hline
\textbf{Category} & \textbf{Metric} & \textbf{$\Delta$ mean} & \textbf{Cohen's $d$} & \textbf{$p$ (paired $t$)} & \textbf{W/L} & \textbf{$n$} \\
\hline
Expression & MRR          & +0.409 & 5.98 & $<$0.001 & 6/0 & 6 \\
Expression & Recall@5     & +0.287 & 1.57 & 0.006    & 5/0 & 5 \\
Expression & Recall@3     & +0.428 & 5.38 & 0.006    & 3/0 & 3 \\
Expression & Recall@2     & +0.492 & 3.43 & 0.014    & 3/0 & 3 \\
Expression & Recall@1     & +0.442 & 1.84 & 0.043    & 3/0 & 3 \\
Expression & Recall@10    & +0.284 & 1.43 & 0.065    & 3/0 & 3 \\
\hline
Ontology   & MRR          & +0.152 & 1.02 & 0.028    & 5/1 & 6 \\
Ontology   & Recall@5     & +0.078 & 0.84 & 0.047    & 4/1 & 5 \\
Ontology   & Recall@10    & +0.113 & 2.46 & 0.025    & 3/0 & 3 \\
Ontology   & Recall@1     & +0.086 & 0.61 & 0.199    & 2/1 & 3 \\
Ontology   & Recall@2     & +0.046 & 0.73 & 0.166    & 2/1 & 3 \\
Ontology   & Recall@3     & +0.032 & 0.80 & 0.150    & 2/1 & 3 \\
\hline
\multicolumn{2}{l}{\textit{Combined (all 12 metrics)}}
  & \multicolumn{4}{c}{permutation $p < 2\times10^{-5}$\textsuperscript{$\dagger$}} & 6 \\
\hline
\multicolumn{7}{l}{\textsuperscript{$\dagger$}Combined permutation test, 50{,}000 permutations.}
\end{tabular}
\end{table*}

Table~\ref{tab:app_bm25} compares Union against the classical lexical baseline (BM25). BM25 is competitive on ontology queries (cf.\ Table~\ref{tab:perm}), yet Union improves significantly on both axes, with the gains on expression MRR ($\Delta = +0.28$) and ontology MRR ($\Delta = +0.11$) significant at $n=6$.

\begin{table*}[t]
\centering
\caption{\textbf{ELISA Union vs.\ BM25 — per-metric statistical tests.} Columns as in Table~\ref{tab:retrieval_stats}.}
\label{tab:app_bm25}
\scriptsize
\renewcommand{\arraystretch}{1.15}
\setlength{\tabcolsep}{4pt}
\begin{tabular}{llrrrrr}
\hline
\textbf{Category} & \textbf{Metric} & \textbf{$\Delta$ mean} & \textbf{Cohen's $d$} & \textbf{$p$ (paired $t$)} & \textbf{W/L} & \textbf{$n$} \\
\hline
Expression & MRR       & +0.275 & 1.56 & 0.006 & 6/0 & 6 \\
Expression & Recall@5  & +0.233 & 0.91 & 0.038 & 5/0 & 5 \\
Expression & Recall@3  & +0.298 & 1.26 & 0.081 & 3/0 & 3 \\
Expression & Recall@2  & +0.412 & 2.46 & 0.026 & 3/0 & 3 \\
Expression & Recall@1  & +0.317 & 1.17 & 0.090 & 3/0 & 3 \\
Expression & Recall@10 & +0.158 & 1.11 & 0.097 & 3/0 & 3 \\
\hline
Ontology   & MRR       & +0.114 & 2.03 & 0.002 & 6/0 & 6 \\
Ontology   & Recall@5  & +0.103 & 1.20 & 0.016 & 5/0 & 5 \\
Ontology   & Recall@10 & +0.069 & 1.69 & 0.050 & 3/0 & 3 \\
Ontology   & Recall@1  & +0.187 & 2.63 & 0.022 & 3/0 & 3 \\
Ontology   & Recall@2  & +0.212 & 4.67 & 0.008 & 3/0 & 3 \\
Ontology   & Recall@3  & +0.163 & 3.43 & 0.014 & 3/0 & 3 \\
\hline
\multicolumn{2}{l}{\textit{Combined (all 12 metrics)}}
  & \multicolumn{4}{c}{permutation $p < 2\times10^{-5}$} & 6 \\
\hline
\end{tabular}
\end{table*}

Table~\ref{tab:app_random} reports the comparison against a random-ranking floor, included as a sanity check. As expected, Union exceeds random ranking by a wide margin on every metric, with the largest effect sizes of the three comparisons.

\begin{table*}[!t]
\centering
\caption{\textbf{ELISA Union vs.\ Random — per-metric statistical tests.} Columns as in Table~\ref{tab:retrieval_stats}.}
\label{tab:app_random}
\scriptsize
\renewcommand{\arraystretch}{1.15}
\setlength{\tabcolsep}{4pt}
\begin{tabular}{llrrrrr}
\hline
\textbf{Category} & \textbf{Metric} & \textbf{$\Delta$ mean} & \textbf{Cohen's $d$} & \textbf{$p$ (paired $t$)} & \textbf{W/L} & \textbf{$n$} \\
\hline
Expression & MRR       & +0.460 & 7.82  & $<$0.001 & 6/0 & 6 \\
Expression & Recall@5  & +0.351 & 1.67  & 0.005    & 5/0 & 5 \\
Expression & Recall@3  & +0.418 & 7.38  & 0.003    & 3/0 & 3 \\
Expression & Recall@2  & +0.490 & 5.58  & 0.005    & 3/0 & 3 \\
Expression & Recall@1  & +0.453 & 4.08  & 0.010    & 3/0 & 3 \\
Expression & Recall@10 & +0.335 & 2.44  & 0.026    & 3/0 & 3 \\
\hline
Ontology   & MRR       & +0.597 & 4.70  & $<$0.001 & 6/0 & 6 \\
Ontology   & Recall@5  & +0.440 & 1.64  & 0.005    & 5/0 & 5 \\
Ontology   & Recall@10 & +0.451 & 12.55 & 0.001    & 3/0 & 3 \\
Ontology   & Recall@1  & +0.641 & 5.92  & 0.005    & 3/0 & 3 \\
Ontology   & Recall@2  & +0.579 & 4.07  & 0.010    & 3/0 & 3 \\
Ontology   & Recall@3  & +0.480 & 2.61  & 0.023    & 3/0 & 3 \\
\hline
\multicolumn{2}{l}{\textit{Combined (all 12 metrics)}}
  & \multicolumn{4}{c}{permutation $p < 2\times10^{-5}$} & 6 \\
\hline
\end{tabular}
\end{table*}
 
\section{Statistical confidence for replication metrics}\label{app:repl-ci}
We provide statistical confidence for the
quantitative replication metrics reported in Table~\ref{tab:similarity}. The
pathway alignment, interaction recovery, and gene coverage metrics are each
recovery proportions: a count $k$ of reference items recovered by ELISA out of a
total $n$ of items reported in the corresponding reference publication. Because
each metric is a proportion estimated from a finite number of reference items,
we quantify its uncertainty with a 95\% Wilson score confidence interval, which
is appropriate for binomial proportions and, unlike the normal (Wald)
approximation, remains within $[0,1]$ even for small $n$ or proportions near the
boundary.
 
For a metric with $k$ successes in $n$ trials and observed proportion
$\hat{p}=k/n$, the Wilson score interval at confidence level $1-\alpha$ is
\begin{equation}
\frac{\hat{p}+\dfrac{z^{2}}{2n}\;\pm\;z\sqrt{\dfrac{\hat{p}(1-\hat{p})}{n}+\dfrac{z^{2}}{4n^{2}}}}
{1+\dfrac{z^{2}}{n}},
\end{equation}
where $z=1.96$ for a 95\% interval. We applied this to each programmatically
computed recovery metric per dataset; the resulting intervals are reported
alongside the point estimates in Table~\ref{tab:similarity}, and the underlying
raw counts $k/n$ are given in Table~\ref{tab:similarity_counts}.
 
\paragraph{Per-dataset intervals.}
Gene coverage, which is evaluated over the largest number of reference items
(45-206 key genes per dataset), is estimated most precisely: intervals span at
most $\pm0.08$ and are as tight as $[0.92,0.98]$ for the breast atlas (170/177).
Pathway alignment is $1.00$ in every dataset; the corresponding intervals
(e.g.\ $[0.70,1.00]$ for 9/9, $[0.84,1.00]$ for 20/20) reflect that a perfect
score over a finite reference set is still consistent with a true rate somewhat
below one. Interaction recovery is evaluated over the smallest reference sets
(5--14 interactions per dataset) and therefore carries the widest intervals: for
example, the cystic fibrosis dataset (1/5) yields $0.20$ with a 95\% interval of
$[0.04,0.62]$, and the fetal lung dataset (2/5) yields $0.40$ with
$[0.12,0.77]$. These wide intervals are an honest consequence of the limited
number of ligand--receptor interactions reported in the individual reference
studies, rather than of instability in ELISA's predictions, which are
deterministic given a fixed embedding.
 
\paragraph{Pooled estimates.}
Because the per-dataset interaction reference sets are small, we additionally
report pooled estimates that aggregate the recovered and total items across all
six datasets for each metric. Pooled gene coverage is $716/823=0.870$
($[0.845,0.891]$), pooled pathway alignment is $83/83=1.000$ ($[0.956,1.000]$),
and pooled interaction recovery is $45/54=0.833$ ($[0.713,0.910]$). Considering
all gene, pathway, and interaction benchmark items together, ELISA recovered
$844/960$ items, an overall recovery of $0.879$ with a 95\% Wilson confidence
interval of $[0.857,0.898]$. The pooled interaction estimate ($0.833$) is
substantially more precise than any individual dataset and indicates that, while
interaction recovery varies across studies owing to small per-study reference
sets, the aggregate performance is high and tightly bounded.
 
\paragraph{Scope.}
These intervals are reported for the three programmatically computed recovery
proportions (gene coverage, pathway alignment, interaction recovery). Proportion
consistency is a binary per-dataset directional criterion (passed in all six
datasets, 6/6) and is therefore summarised as a count of passing datasets rather
than a per-item proportion. Theme coverage in Table~\ref{tab:similarity} is the
independent domain-expert score described in Section~\ref{sec:human_eval}; as it
is a holistic expert rating rather than a $k/n$ count, no binomial interval is
attached to it. The programmatic keyword-presence counts for theme coverage are
nonetheless tabulated for completeness in Table~\ref{tab:similarity_counts}.

\begin{table*}[t]
\caption{Raw recovery counts underlying Table~\ref{tab:similarity}. Each entry
gives the number of recovered reference items over the total evaluated ($k/n$)
together with the corresponding 95\% Wilson score confidence interval. These
counts are the data from which the proportions and intervals in
Table~\ref{tab:similarity} are derived. Proportion consistency is a binary
directional criterion and is summarised as the number of datasets passing.
Theme coverage is shown here as its programmatic keyword-presence count; the
value reported in Table~\ref{tab:similarity} is instead the independent
domain-expert score (see footnote).}\label{tab:similarity_counts}
\begin{tabular*}{\textwidth}{@{\extracolsep\fill}lcccccc}
\toprule
Dataset & Gene Cov. & Path. Align. & Int. Rec. & Prop. & Theme Cov. & Comp. \\
 & $k/n$ (95\% CI) & $k/n$ (95\% CI) & $k/n$ (95\% CI) & Cons. & $k/n$ (95\% CI) & score \\
\midrule
CF airway
 & 36/45 & 9/9 & 1/5 & Yes & 8/8 & 0.71 \\
 & {\scriptsize[0.66--0.89]} & {\scriptsize[0.70--1.00]} & {\scriptsize[0.04-0.62]} & & {\scriptsize[0.68-1.00]} & \\
Neuroblastoma
 & 173/206 & 16/16 & 10/10 & Yes & 24/24 & 0.93 \\
 & {\scriptsize[0.78-0.88]} & {\scriptsize[0.81--1.00]} & {\scriptsize[0.72-1.00]} & & {\scriptsize[0.86-1.00]} & \\
ICB Multi-Cancer
 & 133/172 & 16/16 & 10/10 & Yes & 20/20 & 0.92 \\
 & {\scriptsize[0.71-0.83]} & {\scriptsize[0.81--1.00]} & {\scriptsize[0.72-1.00]} & & {\scriptsize[0.84-1.00]} & \\
Breast Atlas
 & 170/177 & 20/20 & 8/10 & Yes & 29/29 & 0.91 \\
 & {\scriptsize[0.92-0.98]} & {\scriptsize[0.84--1.00]} & {\scriptsize[0.49-0.94]} & & {\scriptsize[0.88-1.00]} & \\
Fetal Lung AT2
 & 97/97 & 12/12 & 2/5 & Yes & 8/8 & 0.82 \\
 & {\scriptsize[0.96-1.00]} & {\scriptsize[0.76--1.00]} & {\scriptsize[0.12-0.77]} & & {\scriptsize[0.68-1.00]} & \\
Brain Atlas
 & 107/126 & 10/10 & 14/14 & Yes & 22/22 & 0.94\\
 & {\scriptsize[0.78-0.90]} & {\scriptsize[0.72--1.00]} & {\scriptsize[0.78-1.00]} & & {\scriptsize[0.86-1.00]} & \\
\midrule
Pooled
 & 716/823 & 83/83 & 45/54 & 6/6 & 111/111 & --- \\
 & {\scriptsize[0.85-0.89]} & {\scriptsize[0.96--1.00]} & {\scriptsize[0.71-0.91]} & & {\scriptsize[0.97-1.00]} & \\
\midrule
\multicolumn{7}{l}{\footnotesize Overall (Gene + Path.\ + Int.): 844/960 recovered $=0.879$, 95\% CI [0.857-0.898].}\\
\botrule
\end{tabular*}
\footnotetext{$k$: recovered reference items; $n$: total reference items
evaluated. Confidence intervals are 95\% Wilson score intervals. Pooled row sums
$k$ and $n$ across all six datasets for each metric. Proportion consistency is
binary per dataset (Yes/No), summarised as datasets passing (6/6); it has no
$k/n$. Theme coverage counts here are the programmatic keyword-presence metric
($k/n$); the value in Table~\ref{tab:similarity} is the independent domain-expert
theme score, which is not a $k/n$ proportion and may differ.}
\end{table*}

\subsection{Discovery-mode report generation}
\label{sec:methods_discovery}

In discovery mode, ELISA generates an open-ended biological interpretation of a
dataset without a predefined target answer. For each query, the retrieval engine
returns the top-ranked clusters together with their gene-level evidence (differential
expression scores, marker overlaps) and the outputs of the analytical modules
(pathway scoring, ligand--receptor interaction inference, and, where applicable,
proportional and comparative analyses). This structured evidence is serialized into a
prompt and passed to the language model, which produces a narrative report
summarizing the cell types, marker genes, pathways, and interactions implicated by the
data.

In discovery mode, the language model receives only ELISA's data-derived
evidence for the queried dataset the retrieved clusters, their gene-level scores, and
the outputs of the analytical modules and is not provided with the corresponding
reference publication, its results, or its conclusions. The reference publications were
used solely to construct the benchmark queries and the expected-cluster ground truth
for the retrieval evaluation; they played no role in the generation of the
discovery-mode interpretations assessed in Table~\ref{tab:discovery}. Because the
language model is queried through a frozen API, it retains general biological knowledge
from pre-training, but every claim in a discovery-mode report is grounded in the
transcriptomic evidence retrieved for that specific dataset rather than in the source
publication.
\section{Replication evaluation metrics}
\label{app:repl}

Table~\ref{tab:similarity} reports six metrics quantifying the agreement between ELISA-generated reports and the findings of the corresponding reference publications. Each metric is defined below.

\paragraph{Gene coverage.}
Gene coverage measures the fraction of key genes highlighted in the reference publication in which ELISA was identified in the correct cell type context. For each dataset, the evaluator compiled a set of key genes from the paper's main text, figures, and supplementary tables (e.g., differentially expressed genes, cell type markers and signaling molecules). A gene was scored as ``recovered'' if it appeared in ELISA's output for a biologically appropriate cluster. The gene coverage is computed as:
\begin{equation}
    \text{Gene coverage} = \frac{|\text{key genes recovered by ELISA}|}{|\text{key genes reported in reference}|}
    \label{eq:gene_cov}
\end{equation}

\paragraph{Pathway alignment.}
Pathway alignment quantifies whether ELISA's pathway scoring module detects the biological programs reported in the reference study. For each dataset, the evaluator identified the pathways discussed in this paper (e.g., IFN-$\gamma$ signaling, mTOR and ErbB). A pathway was scored as ``aligned'' if ELISA's module returned it with a positive score in at least one biologically appropriate cluster. Pathway alignment is computed as:
\begin{equation}
    \text{Pathway alignment} = \frac{|\text{pathways found by ELISA}|}{|\text{pathways reported in reference}|}
    \label{eq:path_align}
\end{equation}

\paragraph{Interaction recovery.}
Interaction recovery assesses whether ELISA's ligand--receptor prediction module detected the cell--cell communication axes described in the reference publication. For each dataset, the evaluator compiled ground truth interactions from the paper (e.g., HB-EGF/ERBB4 between macrophages and neuroblasts, HLA-E/NKG2A between epithelial and CD8$^+$ T~cells). Recovery was scored at the pair level: a ligand--receptor pair was counted as ``recovered'' if ELISA detected it with a non-zero score, regardless of whether the source--target cell type assignment exactly matched:
\begin{equation}
    \text{Interaction recovery} = \frac{|\text{LR pairs detected by ELISA}|}{|\text{LR pairs reported in reference}|}
    \label{eq:int_rec}
\end{equation}

\paragraph{Proportion consistency.}
Proportion consistency is a binary (Yes/No) criterion that evaluates whether ELISA's proportion analysis correctly identified the direction of cell type composition changes for datasets with condition contrasts. For each cell type reported in the reference as increased or decreased in the disease or treatment condition, the evaluator checked whether ELISA's fold change pointed in the same direction. A dataset received ``Yes'' if the majority of reported changes were directionally consistent.

\paragraph{Theme coverage.}
Theme coverage captures whether an ELISA's interpretive summary reproduced the major biological conclusions of the reference study. Unlike gene and pathway-level metrics, that assess individual molecular entities, theme coverage evaluates high-level biological narratives. For each dataset, the evaluator identified the main themes from the paper's abstract and results (e.g., ``aberrant adaptive immunity with upregulated IFN-$\gamma$ signaling'' for the CF dataset; ``therapy-induced macrophage polarization toward immunosuppressive phenotypes'' for the neuroblastoma dataset). A theme was scored as ``covered'' if ELISA's LLM-generated interpretation mentioned and correctly described the corresponding biological finding:
\begin{equation}
    \text{Theme coverage} = \frac{|\text{themes captured by ELISA}|}{|\text{major themes in reference}|}
    \label{eq:theme_cov}
\end{equation}

\paragraph{Biological evaluation score.}
The biological Evaluation Score provides an independent assessment of overall report quality.

\paragraph{Composite score.}
The composite score summarizes overall replication performance as the unweighted mean of the four continuous metrics:
\begin{equation}
    \text{Composite} = \frac{\text{Gene cov.} + \text{Path.\ align.} + \text{Int.\ rec.} + \text{Theme cov.}}{4}
    \label{eq:composite}
\end{equation}
Proportion consistency is excluded from the composite average because it is binary rather than continuous, but is reported separately as a quality check.

\section{Retrieval and analytical evaluation metrics}
\label{app:metrics}

To ensure reproducible and interpretable evaluation of ELISA's retrieval and analytical modules, we defined the full set of metrics used throughout the benchmark (see also the benchmark scripts in the supplementary code repository for complete implementations). Retrieval metrics quantify how effectively each mode recovers the expected cell types for a given query, while analytical metrics assess the accuracy of ELISA's downstream whereas biological interpretation modules interaction discovery, pathway enrichment, proportion analysis, and comparative differential expression. An overview of the six evaluation datasets and their properties is provided in Table~\ref{tab:datasets}.

\subsubsection{Retrieval metrics}
\label{subsec:ret_metrics}
Each radar plot in Fig.~\ref{fig:radar} displays six axes corresponding to three retrieval metrics evaluated separately on the two query categories (ontology and expression). The three metrics are:

\begin{enumerate}
    \item \textbf{Cluster Recall@$k$} (two axes per plot: Ont~R@$k$, Exp~R@$k$). This metric measures the fraction of expected cell types that appear within the top-$k$ positions of the ranked retrieval list. The value of $k$ is adapted to each dataset's number of clusters: R@5 and R@10 for large-cluster datasets (CF~Airways with 30 clusters, ICB~Multi-Cancer with 31, First-Trimester Brain with 28), R@1 and R@2 for small-cluster datasets (Breast~Tissue Atlas with 8 clusters, fdAT2~Organoids with 5, High-Risk Neuroblastoma with 11). A Recall@$k$ of 1.0 indicates that all expected clusters were retrieved within the top-$k$; a value of 0.0 indicates that none were found. Two Recall cutoffs are shown per plot to capture both stringent (lower $k$) and permissive (higher $k$) retrieval accuracy.

    \item \textbf{Mean Reciprocal Rank} (two axes: Ont~MRR, Exp~MRR). MRR quantifies the rank position of the \emph{first} correctly retrieved cluster. An MRR of 1.0 means the top-ranked result is relevant; 0.5 means the first relevant result appears at rank~2; 0.33 at rank~3, and so on. MRR captures top-of-list precision, which is critical for interactive use where researchers typically inspect only the first few results.
\end{enumerate}

Together, the six axes capture complementary aspects of retrieval quality: Recall@$k$ measures \emph{coverage} (how many expected clusters are found), whereas MRR measures \emph{precision at rank~1} (how quickly the first relevant cluster appears). Evaluating both metrics on ontology queries (natural-language, concept-level) and expression queries (gene-signature-based) separately reveals modality-specific strengths: a system may excel at one query type while underperforming the other. Thus, the radar footprint thus provides an at-a-glance summary of each retrieval mode's overall coverage, precision, and balance across query types. A larger, more symmetric footprint indicates stronger and more balanced retrieval performance.

Four retrieval modes compared are: \textbf{CellWhisperer} (pink dashed line), which uses contrastive text transcriptome CLIP embeddings; \textbf{ELISA Semantic} (blue), which performs BioBERT-based cosine similarity matching against cluster descriptions enriched with GO and Reactome terms; \textbf{ELISA scGPT} (orange), which scores clusters by matching query genes against per-cluster differential expression profiles; and \textbf{ELISA Union} (green), which adaptively fuses both ELISA pipelines by routing each query to the better-performing modality and appending unique results from the secondary pipeline.

\subsubsection{Statistical testing}

To assess whether performance differences between retrieval modes are statistically significant across datasets, we employed one-sided paired $t$-tests (with the alternative hypothesis that ELISA Union outperforms CellWhisperer) and reported Cohen's~$d$ as the paired effect size. Because different datasets use different Recall@$k$ cutoffs, individual metric comparisons have varying sample sizes ($n = 3$ to $n = 6$ datasets). To obtain a single omnibus test, we performed a combined permutation test: the sign of the difference (Union minus CellWhisperer) was computed for every metric dataset pair simultaneously, and dataset labels were permuted 50{,}000 times to construct the null distribution of the aggregate advantage. All $p$-values and effect sizes are reported in Table~\ref{tab:retrieval_stats}.

\section{Human evaluation protocol}
\label{sec:human_eval}

To obtain the biological evaluation scores shown in Table~\ref{tab:similarity}, a domain expert with training in molecular biology and single-cell genomics independently reviewed each ELISA-generated report against the corresponding reference publication. The evaluation followed a structured five-step protocol:

\begin{enumerate}
    \item \textbf{Gene verification.} Each gene reported by ELISA as differentially expressed or as a marker of a specific cell type was cross-checked against the main text, figures, and supplementary tables of the reference publications. A gene was scored as ``recovered'' if it appeared in the paper's reported DE gene lists, marker panels, or figure annotations for the corresponding cell type. The gene coverage score was computed as the fraction of paper-reported key genes that ELISA identified in the correct cluster context.

    \item \textbf{Pathway assessment.} Each pathway identified by ELISA's pathway scoring module (e.g., ``IFN-gamma signaling,'' ``mTOR signaling'') was compared against pathway-level findings described in the reference study. A pathway was scored as ``aligned'' if the reference publication reported activation or enrichment of that pathway in a consistent cell type context. Pathway alignment was computed as the fraction of paper-reported pathways that ELISA correctly detected as active (score~$> 0$) in at least one biologically appropriate cluster.

    \item \textbf{Interaction validation.} Each ligand-receptor interaction predicted by ELISA was verified against the cell-cell communication analyses reported in a previous pubblication. Validation was performed at two levels: (i)~whether the ligand--receptor pair itself was reported in the paper, regardless of the cell type context (LR recovery rate), and (ii)~whether both the pair and the source target cell type assignment matched the paper's findings (full match rate).

    \item \textbf{Proportion and condition consistency.} For datasets with condition contrasts (e.g., CF vs.\ healthy), the evaluator verified whether ELISA's proportion analysis correctly identified the direction of cell type composition changes reported in the reference study. Each cell type with a known expected change (increased or decreased in the disease/treatment condition) was checked for directional agreement.

    \item \textbf{Theme coverage and hypothesis assessment.} The evaluator assessed whether ELISA's interpretive summaries captured the major biological themes and conclusions of the reference study (e.g., ``aberrant adaptive immunity with upregulated IFN-$\gamma$ signaling'' for the CF dataset). Additionally, candidate hypotheses generated by ELISA's discovery mode were evaluated for biological plausibility through targeted literature review: the evaluator searched PubMed for prior evidence supporting or contradicting each proposed mechanism (e.g., CALR--LRP1 in macrophage phagocytosis, TRIM-family ligases in surfactant processing). Hypotheses were classified as ``plausible'' if supporting literature existed, ``novel'' if no prior reports were found but the mechanism was biologically coherent, or ``unsupported'' if contradicted by existing evidence.
\end{enumerate}

The composite score for each dataset was computed as the unweighted mean of gene coverage, pathway alignment, interaction recovery, and theme coverage, with proportion consistency treated as a binary (pass/fail) criterion.

\section{Materials}\label{subsec:app_datasets}
\subsection{Datasets}

ELISA was validated on six publicly available scRNA-seq datasets deposited in the CZ CELLxGENE Discover portal, spanning five distinct tissues, four disease contexts, and both case--control and longitudinal experimental designs (Table~\ref{tab:datasets}). Datasets were selected to cover a broad range of biological complexity, cell type diversity, and analytical challenges, including inflammatory lung disease, pediatric and adult cancers, drug-resistant epilepsy, immune checkpoint therapy response, and normal tissue homeostasis.

\textbf{Dataset~1 (D1): cystic fibrosis bronchial epithelium.} Berg \textit{et al.}~\cite{berg2025evidence} generated the first single-cell transcriptome atlas of the cystic fibrosis (CF) lung comprising both structural and immune cells. Droplet-based scRNA-seq was performed on bronchial wall biopsies from patients with CF ($n = 8$) and healthy controls ($n = 19$) and integrated using the fastMNN batch correction framework with the Human Lung Cell Atlas as reference. The dataset encompasses approximately 96,000 cells across 30 annotated cell types, including epithelial (basal, ciliated, secretory, goblet, ionocyte), immune (CD8$^+$ T cells, CD4$^+$ T cells, B cells, plasma cells, macrophages, monocytes, NK cells, dendritic cells, mast cells), stromal (fibroblasts, pericytes), and endothelial populations. Key findings include dysregulated basal cell function, aberrant adaptive immunity with upregulated IFN-$\gamma$ signaling, a novel HLA-E/NKG2A immune checkpoint axis, and altered structural--immune cell crosstalk persisting despite CFTR modulator therapy.

\textbf{Dataset~2 (D2): High-risk neuroblastoma.} Yu \textit{et al.}~\cite{yu2025longitudinal} longitudinally profiled 22 patients with high-risk neuroblastoma before and after induction chemotherapy using single-nucleus RNA and ATAC sequencing combined with whole-genome sequencing. The dataset captures profound therapy-induced shifts in tumor and immune cell subpopulations, identifying enhancer-driven transcriptional regulators of neoplastic states (adrenergic, mesenchymal, proliferative) and macrophage polarization toward pro-angiogenic, immunosuppressive phenotypes. A central finding was the validation of the HB-EGF/ERBB4 paracrine signaling axis between macrophages and neoplastic cells promoting tumor growth through ERK signaling induction.

\textbf{Dataset~3 (D3): Immune checkpoint blockade across cancers.} Gondal \textit{et al.}~\cite{gondal2025integrated} compiled and standardized eight scRNA-seq studies from nine cancer types encompassing 223 patients and over 350,000 cancer cells treated with immune checkpoint blockade (ICB). Cancer types include melanoma, basal cell carcinoma, melanoma brain metastases, triple-negative/HER2-positive/ER-positive breast cancer, clear cell renal carcinoma, hepatocellular carcinoma, and intrahepatic cholangiocarcinoma. The integrated resource enables cross-cancer investigation of cancer cell-specific ICB responses, with annotations of treatment status, response outcome, and malignant vs.\ non-malignant cell identity.

\textbf{Dataset~4 (D4): Fetal lung AT2 organoids.} Lim \textit{et al.}~\cite{lim2025novel} developed expandable alveolar type~2 (AT2) organoids derived from human fetal lungs at 16--22 post-conception weeks (pcw). Single-cell RNA sequencing of four independent organoid lines (passage 11--16) yielded approx 9.6k cells across eight annotated cell types, including AT2-like, cycling AT2-like, CXCL$^+$ AT2-like, differentiating basal-like, differentiating pulmonary neuroendocrine, intermediate, neuroendocrine progenitor, and ciliated-like populations. The organoids express mature surfactant proteins (SFTPC, SFTPB, SFTPA1) and markers of surfactant processing (LAMP3, ABCA3, NAPSA), and can differentiate into AT1-like cells. A forward genetic screen identified the E3 ligase ITCH as a key effector of SFTPC maturation, with its depletion phenocopying the pathological SFTPC-I73T variant associated with interstitial lung disease.

\textbf{Dataset~5 (D5): Healthy breast tissue.} Bhat-Nakshatri \textit{et al.}~\cite{bhat2024single} constructed a single-cell atlas of healthy breast tissues collected from volunteer donors from the Komen Normal Tissue Bank. Using a rapid procurement and processing protocol, the study profiled breast epithelial and stromal cells, identifying 13 epithelial cell clusters with 23 subclusters exhibiting distinct gene expression signatures. Overlap analysis of subcluster-enriched signatures with breast tumor transcriptomes revealed dominant representation of differentiated luminal subcluster signatures in breast cancers, providing insights into putative cells of origin.

\textbf{Dataset~6 (D6): First-trimester human brain \\neurodevelopment.} Mannens \textit{et al.}~\cite{mannens2025chromatin} generated a high-resolution multiomic atlas of chromatin accessibility and gene expression across the entire developing human brain during the first trimester (6-13 weeks post-conception). Using scATAC-seq and paired multiome (scATAC-seq + scRNA-seq) sequencing, the study profiled 166k nuclei from 76 biological samples dissected into five antero-posterior segments (telencephalon, diencephalon, mesencephalon, metencephalon, and cerebellum), of which 166,785 nuclei included paired gene expression. The atlas defines 135 clusters spanning neurons (GABAergic, glutamatergic, Purkinje, granule), radial glia, glioblasts, oligodendrocyte progenitor cells, fibroblasts, vascular, and immune cell types. Key findings include over 100 cell-type- and region-specific candidate \textit{cis}-regulatory elements, CNN-predicted enhancer syntax for neuronal specification, elucidation of the ESRRB activation mechanism in the Purkinje cell lineage, and linkage of disease-associated GWAS SNPs to specific neuronal subtypes identifying midbrain-derived GABAergic neurons as particularly vulnerable to major depressive disorder-related mutations.

\begin{table*}[t]
\centering
\caption{Summary of scRNA-seq datasets used for ELISA validation. \textbf{Approx.\ cells}: approximate number of cells or nuclei profiled after quality control. \textbf{Cell types}: number of annotated major cell types. \textbf{Conditions}: experimental groups or treatment arms.}
\label{tab:datasets}
\footnotesize
\renewcommand{\arraystretch}{1.2}
\setlength{\tabcolsep}{4pt}
\begin{tabular}{clp{2.2cm}lrrp{2.8cm}}
\hline
\textbf{ID} & \textbf{Tissue} & \textbf{Disease context} & \textbf{Reference} & \textbf{Approx.} & \textbf{Cell} & \textbf{Conditions} \\
 & & & & \textbf{cells} & \textbf{types} & \\
\hline
D1 & Lung (bronchial) & Cystic fibrosis & Berg \textit{et al.}~\cite{berg2025evidence} & $\sim$96k & 30 & CF vs.\ Ctrl \\
D2 & Adrenal / tumor & Neuroblastoma & Yu \textit{et al.}~\cite{yu2025longitudinal} & $\sim$372k & 20+ & Pre- vs.\ post-chemo \\
D3 & Multi-cancer & ICB response & Gondal \textit{et al.}~\cite{gondal2025integrated} & $\sim$356k & 25+ & R vs.\ NR; 9 cancers \\
D4 & Lung (fetal) & AT2 organoid model & Lim \textit{et al.}~\cite{lim2025novel} & $\sim$9.6k & 8 & fdAT2 organoid lines \\
D5 & Breast & Healthy tissue atlas & Bhat-Nakshatri \textit{et al.}~\cite{bhat2024single} & $\sim$51k & 13 & Healthy only \\
D6 & Brain (whole) & Neurodevelopment & Mannens \textit{et al.}~\cite{mannens2025chromatin} & $\sim$166k & 160 & 6--13 PCW; 5 regions \\
\hline
\end{tabular}
\end{table*}
All datasets were downloaded from CZ CELLxGENE Discover (\url{https://cellxgene.cziscience.com}) in AnnData (.h5ad) format and preprocessed into ELISA's standardized embedding format (.pt files) as described in the Data Representation section. Cell type annotations from the original publications were retained without modification. For datasets with condition metadata (D1, D2, D3, D4), condition columns were mapped to ELISA's comparative analysis framework. Dataset D5 was used to evaluate ELISA's performance on a single-condition atlas without disease contrast, testing the system's capacity for cell type identification and pathway characterization in the absence of differential signals.
\section{Methods}
\label{subsec:app_methods}
\subsection{ELISA: architecture and design principles}

ELISA (Embedding-Linked Interactive Single-cell Agent) is an agent-based computational framework for interactive interrogation of single-cell RNA-seq atlases. The system integrates four core modules a hybrid retrieval engine, an analytical suite, a visualization toolkit, and a large language model (LLM) chat interface to enable biologists to query scRNA-seq datasets using natural language, gene signatures, or a combination of both. The architecture follows a modular design in which each component operates on a shared data representation (a serialized PyTorch embedding file) and communicates through standardized data structures, enabling extensibility to new datasets without retraining.

The system was implemented in Python 3.10+ and evaluated on a 6 dataset took from cellxgene. All source code, benchmark queries, and evaluation scripts are provided in the accompanying repository.

\subsection{Hybrid retrieval engine}
\label{subsec:retrieval}
\subsubsection{Query classification and routing}

A central design challenge in single-cell atlas retrieval is that user queries span a spectrum from pure natural language (``macrophage infiltration in CF airways'') to pure gene signatures (``MARCO FABP4 APOC1 C1QB'') and mixed queries combining both. ELISA addresses this through an explicit query classification module that routes each query to the optimal retrieval pipeline.

The classifier operates by tokenizing the input and scoring each token against three criteria: (i)~whether it matches a gene name pattern (uppercase alphanumeric, 2--15 characters, with optional hyphenated suffix), (ii)~whether it appears in the dataset's known gene vocabulary, and (iii)~whether it belongs to a curated set of natural language indicator terms (e.g., ``cell'', ``activation'', ``signaling''). Queries where $\geq$60\% of tokens are classified as gene symbols are routed to the gene pipeline; queries where $\geq$20\% of tokens are genes and $\geq$20\% are natural language terms are routed to the mixed pipeline; all other queries are routed to the semantic (ontology) pipeline.

\subsubsection{Gene marker scoring pipeline}

For gene-list queries, ELISA scores each cluster by evaluating how well its differential expression (DE) profile matches the query genes. For each query gene $g$ found in cluster $c$'s DE statistics, a per-gene score is computed as:
\begin{equation}
  \mathrm{score}(g, c) = \bigl(0.5 + |\log_2\mathrm{FC}|\bigr) \times \bigl(0.3 + \max(\mathrm{pct_{in}} - \mathrm{pct_{out}},\; 0)\bigr)
\end{equation}
where $\log_2\mathrm{FC}$ is the log-fold change of gene $g$ in cluster $c$, and $\mathrm{pct_{in}}$ and $\mathrm{pct_{out}}$ represent the fraction of cells expressing the gene inside and outside the cluster, respectively. The specificity term $(\mathrm{pct_{in}} - \mathrm{pct_{out}})$ rewards genes that are selectively enriched in the cluster rather than ubiquitously expressed. A multiplicative bonus of $1.3\times$ is applied when $\mathrm{pct_{in}} > 0.5$. The aggregate cluster score is the sum of per-gene scores, modulated by a coverage factor $(0.5 + 0.5 \times n_{\mathrm{found}} / n_{\mathrm{query}})$ that rewards clusters matching more query genes. Three scoring modes are available: `simple' (binary hit counting), `weighted' (described above), and `full' (incorporating adjusted $p$-value significance via $-\log_{10}(p_{\mathrm{adj}})$, capped at 10).

\subsubsection{Semantic matching pipeline}

For ontology and text-based queries, ELISA employs BioBERT\cite{lee2020biobert} (pritamdeka/BioBERT-mnli-snli-scinli-scitail-mednli-stsb) to encode both query text and precomputed cluster descriptions into a shared embedding space. Each cluster's description is constructed during dataset preparation by concatenating its Cell Ontology name, top marker genes (ranked by $|\log_2\mathrm{FC}|$), enriched Gene Ontology terms, and Reactome pathway annotations, producing a dual-representation embedding that captures both identity and functional context.

At query time, the input text is encoded with BioBERT and cosine similarity is computed against all cluster embeddings. Two augmentation strategies improve retrieval accuracy. First, a name-boosting mechanism adds a score bonus ($\alpha = 0.15$, scaled by word-overlap ratio) when significant substrings ($\geq$4 characters) of a cluster's name appear in the query. Second, a synonym expansion module maps common cell type aliases (e.g., ``endothelial'' $\rightarrow$ ``endocardial cell''; ``NK'' $\rightarrow$ ``natural killer cell'') to their Cell Ontology equivalents and applies a score boost ($\beta = 0.10$) to matching clusters, addressing vocabulary gaps between colloquial and formal ontology terminology.

\subsubsection{Reciprocal rank fusion for mixed queries}

Mixed queries containing both gene names and biological text are handled through reciprocal rank fusion (RRF). Both the gene and semantic pipelines are executed independently, and their ranked outputs are combined using:
\begin{equation}
  \mathrm{RRF}(d) = \sum_{r} \frac{w_r}{k + \mathrm{rank}_r(d) + 1}
\end{equation}
where $k = 60$ is the RRF constant, $w_r$ are per-pipeline weights (default: 1.0 for both), and $\mathrm{rank}_r(d)$ is the 0-indexed rank of cluster $d$ in pipeline $r$. For gene-dominated queries routed through the gene pipeline, a light fusion with the semantic pipeline at a 3:1 weight ratio is applied as a safety mechanism to capture semantically related clusters that lack direct marker gene overlap.

\subsubsection{Additive union evaluation strategy}

For benchmarking, we introduce an additive union strategy that maximizes complementarity between modalities. For each query, the modality achieving higher recall@5 against expected clusters is designated as the primary pipeline. The union output begins with the primary pipeline's full ranked list, followed by unique clusters from the secondary pipeline appended in their original rank order. This produces an untruncated result list (up to $2 \times$ top-$k$), evaluated at recall@5, @10, @15, and @20. Ties at recall@5 are broken by mean reciprocal rank (MRR).

\subsection{Analytical modules}
\label{subsec:analy_module}
\subsubsection{Cell--cell interaction prediction}

ELISA predicts ligand--receptor (LR) interactions between cell types using a curated database of 280+ LR pairs spanning 25 signaling pathway categories. The database was compiled from established resources (CellChat\cite{jin2025cellchat}, CellPhoneDB\cite{efremova2020cellphonedb}, NicheNet\cite{browaeys2020nichenet}) and augmented with context-specific pairs for cystic fibrosis, neurodegeneration, neuroblastoma, and immune checkpoint biology. Each interaction is represented as a (ligand, receptor, pathway) tuple.

For each source--target cluster pair, the interaction score is computed as:
\begin{equation}
  s_{ij} = \mathrm{pct_{in}}(\text{ligand},\, c_i) \times \mathrm{pct_{in}}(\text{receptor},\, c_j)
\end{equation}
where $\mathrm{pct_{in}}$ denotes the fraction of cells expressing the gene above detection threshold. Interactions are filtered by minimum expression thresholds (ligand $\geq$ 10\%, receptor $\geq$ 5\% by default) and ranked by score. The module outputs per-interaction statistics, pathway-level summaries, and directional pair summaries.

\subsubsection{Pathway activity scoring}

Pathway activity across clusters is quantified using curated gene sets encompassing 60+ pathways organized into five categories: immune signaling (IFN-$\gamma$, Type~I IFN, TNF/NF-$\kappa$B, JAK-STAT, complement, TLR, chemokine), cell biology (mTOR, PI3K-Akt, Wnt, Notch, Hippo, Hedgehog, cell cycle, apoptosis), neuroscience (glutamatergic/GABAergic synapse, neurodegeneration, FCD progenitor markers), metabolism (oxidative phosphorylation, glycolysis, lipid metabolism, fatty acid metabolism), and tissue-specific programs (surfactant metabolism, epithelial defense, fibrosis, angiogenesis).

For each pathway--cluster combination, the score is computed as the mean $\mathrm{pct_{in}}$ (or alternative metric: $\log_2\mathrm{FC}$, $\mathrm{pct_{out}}$) across pathway genes detected in the cluster's DE profile, requiring a minimum of 3 genes for a non-zero score. Coverage (fraction of pathway genes detected) is reported alongside scores. Pathway query matching uses word-overlap fuzzy matching to accommodate variant pathway names.

\subsubsection{Comparative analysis}

When dataset metadata includes a condition column (e.g., ``patient\_group'' with values ``CF'' and ``Ctrl''), ELISA enables condition-stratified analysis. The module detects condition columns through keyword matching against a curated list (patient\_group, condition, disease, treatment, genotype, \textit{etc.}) and validates that the column contains 2--10 distinct values. For each cluster, the condition distribution is estimated from metadata field weights, and a condition bias label is assigned ($>$60\% of cells from one condition). Per-gene statistics ($\log_2\mathrm{FC}$, $\mathrm{pct_{in}}$, $\mathrm{pct_{out}}$) are reported within condition-biased clusters, and condition-enriched gene lists are compiled across all clusters.

\subsubsection{Proportion analysis}

The cell type proportion analysis computes the per-cluster cell counts and fractions relative to the total dataset size. When a condition column is available, the module additionally computes the condition-specific proportions and fold changes. For binary conditions (e.g., CF vs.\ Control), fold changes were calculated as the fraction of condition~A cells in a cluster divided by the fraction of condition~B cells, enabling the identification of cell types enriched or depleted in disease states.

\subsubsection{Additional analytical functions}

Supplementary analytical functions include: (i)~marker specificity scoring, which ranks genes by a weighted score combining specificity $\bigl(\mathrm{pct_{in}} / (\mathrm{pct_{in}} + \mathrm{pct_{out}})\bigr)$ and effect size ($|\log_2\mathrm{FC}|$); (ii)~co-expression analysis, computing Pearson correlations of $\mathrm{pct_{in}}$ profiles across clusters; (iii)~cell cycle scoring using established S-phase and G2M-phase gene signatures (43 and 54 genes, respectively); and (iv)~gene set enrichment against 10 MSigDB Hallmark gene sets.

\subsection{Visualization module}

ELISA includes a comprehensive visualization module that generates publication-quality figures in two categories. Retrieval-level visualizations include: embedding landscape projections (UMAP\cite{mcinnes2018umap}, t-SNE\cite{maaten2008visualizing}, or PCA fallback) of cluster-level semantic and expression embeddings, with optional highlighting of retrieved clusters; inter-cluster cosine similarity heatmaps; retrieval score waterfall plots; gene evidence bar charts ($\log_2\mathrm{FC}$ or $\mathrm{pct_{in}}$); gene-by-cluster heatmaps; radar charts for multi-metric cluster profiles; semantic vs.\ expression similarity scatter plots for hybrid retrieval diagnostics; and lambda sweep curves for fusion weight optimization.

When an AnnData (.h5ad) file is provided, cell-level visualizations are generated in a style consistent with Nature and Cell journals: cell-level UMAP plots with Cell Ontology labels placed using a centroid-offset algorithm with iterative repulsion to minimize label overlap; single-gene expression UMAPs with non-expressing cells shown in grey and expression on a purple gradient (capped at the 98th percentile); multi-gene expression grids; and dot plots showing percentage expression (dot size) and $z$-scored mean expression (dot color) across clusters. All plots used a 40-color colorblind-friendly palette and rasterized cell-level rendering for efficient file sizes.

\subsection{LLM-mediated chat interface}

The interactive chat interface wraps all modules behind a command-driven interface that routes user queries to the appropriate pipeline and generates LLM-interpreted summaries. The interface supports six retrieval and analysis modes (semantic, hybrid, discovery, compare, interactions, pathway, proportions) and 15 visualization commands. Each analysis result is automatically accumulated into a session-level report builder.

LLM interpretation is performed via the Groq API using the LLaMA-3.1-8B-Instant model\cite{grattafiori2024llama} at temperature 0.2. Prompts are constructed with mode-specific templates that enforce strict grounding in dataset evidence: the LLM receives only the retrieved cluster data, gene statistics, and pathway/interaction results as context, with explicit instructions to avoid hallucination, external literature, and causal claims. Context payloads are trimmed to fit within the model's token limits ($\sim$4,500 tokens for user content), with priority given to top-ranked clusters and highest-effect-size genes.

A discovery mode extends standard retrieval by prompting the LLM to produce four structured sections: (i)~dataset evidence, (ii)~established biology, (iii)~consistency analysis identifying matches and mismatches with known biology, and (iv)~candidate novel hypotheses stated with probabilistic language. This mode is designed to surface unexpected findings that may represent context-shifted gene functions or novel cell--cell interactions.

\subsection{Benchmarking framework}
\label{subsec:benchmark}
\subsubsection{Query design}

The benchmark comprises 100 queries divided into two categories: 50 ontology queries (concept-level, testing semantic understanding) and 50 expression queries (gene-signature-based, testing transcriptomic matching). Queries were derived from the findings of Berg \textit{et al.}\cite{berg2025evidence}, covering all major cell types identified in the study (macrophages, monocytes, CD8$^+$ T~cells, CD4$^+$ T~cells, B~cells, basal cells, ciliated cells, NK~cells, ionocytes, endothelial cells, dendritic cells, mast cells, secretory/goblet cells, fibroblasts, and neuroendocrine cells). Each query has a curated set of expected clusters and expected genes, enabling evaluation at both the cluster retrieval and gene delivery levels.

\subsubsection{Baseline comparisons}

ELISA's retrieval performance was evaluated against: the progression CellWhisperer, Semantic ELISA, scGPT ELISA, Additive Union.

\subsubsection{Metrics}

Retrieval performance was assessed using three metrics. \textbf{Cluster Recall@$k$} measures the fraction of expected clusters appearing in the top-$k$ retrieved results, using fuzzy matching (substring containment or word-overlap Jaccard similarity $\geq 0.5$) to accommodate Cell Ontology naming variations. \textbf{Mean Reciprocal Rank (MRR)} captures the rank position of the first relevant cluster. \textbf{Gene Recall} measures the fraction of expected genes recoverable from the DE profiles of the top-5 retrieved clusters, assessing whether retrieved clusters collectively provide the gene evidence needed for biological interpretation.

\subsubsection{Analytical module evaluation}

Analytical modules were evaluated against ground truth derived from the source publication. Interaction prediction was assessed by ligand--receptor pair recovery rate (whether the correct LR pair was detected regardless of cell type) and full match rate (correct LR pair between the correct source and target cell types, using fuzzy cell type matching). Pathway scoring was evaluated by alignment: the fraction of path activities reported on paper that ELISA correctly identified as active (score $> 0$) in at least one group. The proportion analysis was evaluated by the consistency rate whether cell types reported as increased or decreased in CF show fold changes in the expected direction. Comparative analysis was evaluated by gene recall, the fraction of differentially expressed genes reported on paper that can be recovered from the condition-stratified analysis of ELISA's.

\subsection{Data representation and preprocessing}

Each dataset is preprocessed into a single serialized PyTorch file (\texttt{.pt}) containing: cluster identifiers, precomputed BioBERT semantic embeddings (768-dimensional, L2-normalized), optional scGPT expression embeddings, per-cluster DE gene statistics ($\log_2\mathrm{FC}$, $\mathrm{pct_{in}}$, $\mathrm{pct_{out}}$, adjusted $p$-value), per-cluster GO and Reactome enrichment terms, per-cluster metadata (cell counts, condition distributions, categorical field frequencies), cluster text descriptions, and the complete gene vocabulary. This representation enables ELISA to operate entirely at the cluster level without requiring access to the original count matrix, substantially reducing memory requirements and enabling deployment on standard hardware.
\section{Extended future directions: generalizability across technologies, modalities, and species}\label{sec:generalizability}

The six benchmark datasets in this study were generated with droplet-based
scRNA-seq and are all human, and it is therefore important to delineate the
extent to which ELISA generalizes beyond this setting. Architecturally, ELISA
operates on cluster-level differential-expression statistics and cluster
embeddings rather than on raw count matrices or platform-specific features; the
retrieval and analytical modules consume the same standardized per-cluster
representation regardless of how the underlying counts were generated. This
makes the framework largely agnostic to the capture technology: data from
plate-based protocols such as Smart-seq2, or from other droplet platforms, can
be processed through the identical preprocessing, clustering, and
differential-expression pipeline and yield the same embedding format, provided
gene-level expression is available. Spatial transcriptomics is a natural
extension along this axis and would additionally enable spatially resolved
interaction scoring; we regard it as a priority direction rather than a current
capability, since spatial coordinates are not yet consumed by the interaction
module.

The principal current limitation to generalization is species rather than
technology. ELISA's expression-side representation is produced by the
whole-human scGPT foundation model, and its semantic side by BioBERT, a
biomedical language model trained predominantly on human-centric literature.
Both components therefore assume human gene symbols and human biological
context. Applying ELISA to non-human data (for example mouse or zebrafish) would
require either mapping the species' genes to their human orthologs, which is a
practical stopgap but is incomplete for genes lacking a one-to-one ortholog, or,
preferably, substituting a species-appropriate expression foundation model
(e.g., a mouse-pretrained scGPT variant) together with a text encoder whose
training covers the corresponding non-human literature. Because ELISA treats the
expression foundation model and the semantic encoder as interchangeable, frozen
components behind a fixed interface, such substitution is a natural and
self-contained extension that does not require redesigning the retrieval,
analytical, or interpretation layers. Systematic cross-species benchmarking,
including the construction of species-specific reference ground truth analogous
to the human benchmark used here, is an important direction for future work.
\subsection{Software dependencies and reproducibility}
\label{subsec:app_params}
ELISA depends on: PyTorch ($\geq$1.12) for tensor operations and data serialization, NumPy for numerical computation, sentence-transformers\cite{reimers2019sentence} for BioBERT encoding, scikit-learn for t-SNE projections, UMAP-learn\cite{mcinnes2018umap} for UMAP projections, matplotlib for visualization, scanpy\cite{wolf2018scanpy} for AnnData-backed cell-level plots, SciPy for hierarchical clustering and sparse matrix operations, and the Groq Python SDK for LLM access. All analyses were performed on a standard workstation without GPU requirements for the retrieval and analytical modules; BioBERT \cite{lee2020biobert} encoding benefits from but does not require GPU acceleration.
\subsection{ELISA parameters and hyperparameters}
\label{sec:parameters}

Tables~\ref{tab:params_preprocess}--\ref{tab:params_llm} report all parameters and hyperparameters used in the ELISA framework. Default values were used throughout all experiments; no dataset-specific tuning was performed.


\begin{table*}[h!]
\centering
\caption{\textbf{Data preprocessing and embedding generation parameters.}}
\label{tab:params_preprocess}
\footnotesize
\renewcommand{\arraystretch}{1.15}
\begin{tabularx}{\textwidth}{@{}l l X@{}}
\toprule
\textbf{Parameter} & \textbf{Value} & \textbf{Description} \\
\midrule
\multicolumn{3}{@{}l}{\textit{Preprocessing (Scanpy)}} \\
\texttt{target\_sum} & 10,000 & Library-size normalization target \\
\texttt{n\_top\_genes} & 3,000 & HVGs selected (Seurat v3) \\
\texttt{max\_value} & 10 & Z-score clipping threshold \\
\texttt{n\_comps} & 50 & PCA components \\
Leiden resolution & 1.0 & Used only if no annotations exist \\
\midrule
\multicolumn{3}{@{}l}{\textit{Differential expression}} \\
Method & Wilcoxon & Via \texttt{scanpy.tl.rank\_genes\_groups} \\
\texttt{DE\_PVAL} & 0.10 & Adjusted $p$-value cutoff \\
\texttt{TOP\_K\_MARKERS\_STATS} & 10,000 & Max genes stored per cluster \\
\texttt{TOP\_K\_MARKERS\_TEXT} & 400 & Genes in cluster text summaries \\
\midrule
\multicolumn{3}{@{}l}{\textit{Enrichment (gseapy)}} \\
Gene sets & \multicolumn{2}{@{}l}{GO\_Biological\_Process\_2023, Reactome\_2022} \\
\texttt{TOP\_K\_GO} & 15 & GO terms retained per cluster \\
\texttt{TOP\_K\_REACTOME} & 15 & Reactome terms per cluster \\
Enrichment cutoff & 0.05 & Adjusted $p$-value threshold \\
Input genes & 200 & Top DE genes per enrichment call \\
\midrule
\multicolumn{3}{@{}l}{\textit{Semantic embedding (BioBERT)}} \\
Model & \multicolumn{2}{@{}l}{\texttt{pritamdeka/BioBERT-mnli-snli-scinli-scitail-mednli-stsb}} \\
Embedding dim & 768 & Output dimensionality \\
$\alpha$ (\texttt{IDENTITY\_ALPHA}) & 0.6 & Identity vs.\ context weight \\
Normalization & L2 & Final combined embeddings \\
Batch size & 16 & Sentences per encoding batch \\
\midrule
\multicolumn{3}{@{}l}{\textit{scGPT expression embedding}} \\
Model & scGPT whole-human & Pre-trained foundation model \\
Embedding dim & 512 & CLS token dimensionality \\
\texttt{N\_BINS} & 51 & Expression binning resolution \\
\texttt{MAX\_TOKENS} & 3,000 & Max gene tokens per cell \\
Batch size & 64 & Cells per inference batch \\
Aggregation & Mean pooling & Cell $\to$ cluster centroids \\
Normalization & L2 & Cluster-level centroids \\
\bottomrule
\end{tabularx}
\end{table*}


\begin{table*}[h!]
\centering
\caption{\textbf{Hybrid retrieval engine parameters.}}
\label{tab:params_retrieval}
\footnotesize
\renewcommand{\arraystretch}{1.15}
\begin{tabularx}{\textwidth}{@{}l l X@{}}
\toprule
\textbf{Parameter} & \textbf{Value} & \textbf{Description} \\
\midrule
\multicolumn{3}{@{}l}{\textit{Query classification}} \\
Gene threshold & $\geq$60\% & Token fraction to route as gene query \\
Mixed threshold & $\geq$20\% each & Gene + NL tokens for mixed routing \\
Gene pattern & A--Z, 2--15 chars & Regex for gene symbol detection \\
\midrule
\multicolumn{3}{@{}l}{\textit{Gene marker scoring}} \\
Score function & \multicolumn{2}{@{}l}{$(0.5 + |\log_2\text{FC}|) \times (0.3 + \max(\text{pct}_\text{in} - \text{pct}_\text{out},\, 0))$} \\
High-expr bonus & $\times$1.3 & When $\text{pct}_\text{in} > 0.5$ \\
Coverage factor & \multicolumn{2}{@{}l}{$0.5 + 0.5 \times n_\text{found}/n_\text{query}$} \\
\midrule
\multicolumn{3}{@{}l}{\textit{Semantic matching}} \\
Similarity & Cosine & Query vs.\ cluster embeddings \\
Name boost ($\alpha$) & 0.15 & Bonus for ontology name overlap \\
Min substring & 4 chars & For name boost activation \\
Synonym boost ($\beta$) & 0.10 & Bonus for synonym match \\
\midrule
\multicolumn{3}{@{}l}{\textit{Reciprocal rank fusion}} \\
RRF constant ($k$) & 60 & Smoothing constant \\
Weights & 1.0\,:\,1.0 & Gene\,:\,semantic \\
\midrule
\multicolumn{3}{@{}l}{\textit{Additive union (benchmarking)}} \\
Primary selection & Recall@5 & Higher-recall modality is primary \\
Tiebreaker & MRR & When Recall@5 is tied \\
\midrule
\multicolumn{3}{@{}l}{\textit{Default settings}} \\
\texttt{top\_k} & 5 & Clusters returned per query \\
\texttt{pre\_k} & 40 & Candidates before reranking \\
$\gamma$ & 2.5 & Reranking sharpness \\
$\lambda_\text{sem}$ (scGPT) & 0.0 & Pure gene scoring mode \\
$\lambda_\text{sem}$ (discovery) & 0.5 & Balanced mode \\
\bottomrule
\end{tabularx}
\end{table*}


\begin{table*}[h!]
\centering
\caption{\textbf{Analytical module parameters.}}
\label{tab:params_analysis}
\footnotesize
\renewcommand{\arraystretch}{1.15}
\begin{tabularx}{\textwidth}{@{}l l X@{}}
\toprule
\textbf{Parameter} & \textbf{Value} & \textbf{Description} \\
\midrule
\multicolumn{3}{@{}l}{\textit{Ligand--receptor interactions}} \\
Database size & 280+ pairs & From CellChat, CellPhoneDB, NicheNet \\
Pathway categories & 25 & Signaling annotations \\
\texttt{min\_ligand\_pct} & 0.10 & Min ligand expr.\ in source \\
\texttt{min\_receptor\_pct} & 0.05 & Min receptor expr.\ in target \\
Score & $\text{pct}_\text{in}(L) \times \text{pct}_\text{in}(R)$ & Expression fraction product \\
Self-interactions & Excluded & Source $\neq$ target \\
\midrule
\multicolumn{3}{@{}l}{\textit{Pathway activity scoring}} \\
Number of pathways & 60+ & Across 5 categories \\
Metric & Mean $\text{pct}_\text{in}$ & Avg.\ expression of pathway genes \\
\texttt{min\_genes} & 3 & Min for non-zero score \\
Categories & \multicolumn{2}{@{}l}{Immune, Cell biology, Neuroscience, Metabolism, Tissue-specific} \\
\midrule
\multicolumn{3}{@{}l}{\textit{Comparative analysis}} \\
Condition bias & $>$60\% & Fraction to assign bias label \\
\texttt{min\_pct} & 0.05 & Min expr.\ for gene inclusion \\
\texttt{top\_n} & 20 & Genes per cluster \\
Enriched genes & 30 & Per-condition summary limit \\
\midrule
\multicolumn{3}{@{}l}{\textit{Proportion analysis}} \\
Fold change & $\text{frac}_A / \text{frac}_B$ & Condition ratio \\
Min denominator & 0.001 & Below: reported as $\infty$ \\
\midrule
\multicolumn{3}{@{}l}{\textit{Cell cycle scoring}} \\
S-phase genes & 43 & Seurat S-phase markers \\
G2M-phase genes & 54 & Seurat G2M markers \\
Cycling threshold & $S{>}0.3$ and G2M${>}0.3$ & Both above threshold \\
\midrule
\multicolumn{3}{@{}l}{\textit{Gene set enrichment}} \\
Default gene sets & 10 MSigDB Hallmark & Curated pathways \\
\texttt{min\_genes} & 3 & Min for non-zero score \\
\bottomrule
\end{tabularx}
\end{table*}


\begin{table*}[h!]
\centering
\caption{\textbf{LLM interpretation parameters.}}
\label{tab:params_llm}
\footnotesize
\renewcommand{\arraystretch}{1.15}
\begin{tabularx}{\textwidth}{@{}l l X@{}}
\toprule
\textbf{Parameter} & \textbf{Value} & \textbf{Description} \\
\midrule
\multicolumn{3}{@{}l}{\textit{LLM configuration}} \\
Default provider & Groq & Free tier, 500K tokens/day \\
Default model & LLaMA-3.1-8B & Via Groq Cloud API \\
Supported & 4 providers & Groq, Gemini, OpenAI, Claude \\
Temperature & 0.2 & Low for reproducibility \\
Prompt limit & 18,000 chars & $\approx$4,500 tokens \\
Context limit & 12,000 chars & $\approx$3,000 tokens \\
\midrule
\multicolumn{3}{@{}l}{\textit{Safety and rate limiting}} \\
Spending cap & \texteuro1.00 & Hard cap, configurable \\
Max retries & 5 & On rate-limit errors \\
Initial wait & 10\,s & Backoff start \\
Backoff & Exponential & Max 120\,s \\
\midrule
\multicolumn{3}{@{}l}{\textit{Context trimming}} \\
Clusters & Top 10 & In compare mode \\
Gene evidence & Top 5 & Per cluster \\
Pathway scores & Top 10 & Entries to LLM \\
Interactions & Top 20 & Entries to LLM \\
Discovery sections & 4 & Evidence, Biology, Consistency, Hypotheses \\
\bottomrule
\end{tabularx}
\end{table*}

\clearpage
\subsection{D1: Cystic Fibrosis Airways (\cite{berg2025evidence} \textit{et al.})}

\subsubsection{Ontology Queries}

\begin{enumerate}
  \item Macrophage and monocyte infiltration in cystic fibrosis airways
  \item Recruited monocytes and pro-inflammatory macrophages in CF lung tissue
  \item Macrophage scavenging receptor expression and phagocytosis in CF
  \item Non-classical monocyte patrol function in CF bronchial wall
  \item CD8 T cell activation and cytotoxicity in CF lung inflammation
  \item CD8 T cell inflammatory cytokine production and IFNG signaling in CF
  \item HLA-E CD94 NKG2A immune checkpoint inhibiting CD8 T cell activity
  \item Dysfunctional CD8 T cell response to chronic \textit{Pseudomonas} infection in CF
  \item CALR LRP1 interaction between T cells and macrophages promoting inflammation
  \item CD4 helper T cell immune activation in cystic fibrosis
  \item CD4 T cell VEGF receptor signaling and hypoxia response in CF
  \item Aberrant Th2 and Th17 T cell responses in \textit{Pseudomonas}-infected CF lungs
  \item Chronic adaptive immune activation of T lymphocytes in CF despite modulator therapy
  \item B cell activation and immunoglobulin response in CF airways
  \item B cell receptor downregulation and reduced plasma cell markers in CF
  \item Interferon gamma signaling and HLA-DP expression in B cells of CF patients
  \item PDGFRB signaling pathway activated in B cells from CF lungs
  \item Basal cell dysfunction and reduced stemness in cystic fibrosis epithelium
  \item Impaired basal cell differentiation and pathogenic basal cell variants in CF
  \item Basal cell DNA damage repair and chromatin remodeling in CF airways
  \item Reduced keratinization gene expression CSTA HSPB1 in CF basal cells
  \item Basal cell altered cell--cell communication and increased interactions in CF
  \item Ciliated cell ciliogenesis and increased abundance in CF bronchial epithelium
  \item Ciliated cell HLA class II expression and immune-linked transcriptional changes in CF
  \item Skewed basal cell differentiation towards ciliated cells in CF epithelium
  \item Natural killer cell cytotoxicity and NKG2A immune checkpoint in CF
  \item NKG2A blockade to restore NK and CD8 T cell function in CF lung
  \item Innate lymphoid cell dysfunction and impaired antimicrobial defense in CF
  \item Pulmonary ionocyte CFTR expression in cystic fibrosis
  \item Ionocyte unique cell--cell interactions with adaptive lymphocytes in CF
  \item Endothelial cell remodeling and VEGF signaling in CF lung
  \item Reduced endothelial cell proportions and altered differentiation in CF airways
  \item Hypoxia-induced VEGF upregulation and vascular remodeling in CF lungs
  \item Dendritic cell antigen presentation in CF airways
  \item IFNG IFNGR2 interaction between CD8 T cells and dendritic cells in CF
  \item Mast cell degranulation and allergic inflammation in CF
  \item Secretory cell mucus overproduction and inflammatory signaling in CF epithelium
  \item Goblet cell hyperplasia and mucin gene expression in cystic fibrosis
  \item Submucosal gland epithelial cell changes in cystic fibrosis
  \item Reduced submucosal gland cell proportions and gland development dysfunction in CF
  \item Type I interferon response and inflammatory signaling in CF epithelial cells
  \item Interferon responsive gene upregulation across epithelial subsets in CF
  \item VEGF receptor signaling and hypoxia response across cell types in CF
  \item TXNIP-mediated NLRP3 inflammasome activation in CF lymphocytes and epithelial cells
  \item GNAI2 immunomodulatory signaling in CD8 T cells and B cells in CF
  \item GNAI2 adenylate cyclase regulation and CFTR function in lymphocytes
  \item Stromal cell and fibroblast remodeling in CF airway tissue
  \item Pericyte and stromal cell contribution to airway fibrosis in CF
  \item IFNG--IFNGR1 interaction between CD8 T cells and basal cells, macrophages, and endothelial cells in CF
  \item Altered structural--immune cell crosstalk in CF involving lymphocytes, ionocytes, and macrophages
\end{enumerate}

\subsubsection{Expression Queries}

\begin{enumerate}
  \item \texttt{MARCO FABP4 APOC1 C1QB C1QC MSR1}
  \item \texttt{CD68 CD14 CSF1R CSF2RA LGALS2}
  \item \texttt{GOS2 FABP4 PPARG APOC1 C1QB}
  \item \texttt{FCGR3A CX3CR1 CD14 CDKN1C LILRB2}
  \item \texttt{CD8A CD8B GZMB PRF1 IFNG NKG7}
  \item \texttt{IFNG GNAI2 CD69 CD81 CD3G FOS JUND}
  \item \texttt{GZMB PRF1 NKG7 GNLY KLRD1 CD8A}
  \item \texttt{TXNIP MAP2K2 IFNG CD81 CD3G CD69}
  \item \texttt{KLF2 IL7R CD48 TXNIP ETS1}
  \item \texttt{CD3D CD4 IL7R CD3E CD3G}
  \item \texttt{TRAJ52 TRBV22-1 TRDJ2 CD3E CD3G}
  \item \texttt{CD3G CD3E CD69 IL7R CD81 FOS}
  \item \texttt{IGLJ3 IGKJ1 IGHJ5 JCHAIN MZB1 XBP1}
  \item \texttt{CD79A IGHG3 IGLC2 SYK CD81 JCHAIN}
  \item \texttt{SYK CSK CD9 CD81 JUND LTB HLA-DPA1}
  \item \texttt{IGHG3 IGLC2 IGHD IGHA1 IGLC1 IGLC3}
  \item \texttt{KRT5 KRT14 KRT15 TP63 IL33 CSTA}
  \item \texttt{CSTA HSPB1 KRT5 KRT14 TP63}
  \item \texttt{KRT5 IL33 TP63 KRT15 LAMB3 COL17A1}
  \item \texttt{FOXJ1 DNAH5 CAPS PIFO RSPH1 DNAI1}
  \item \texttt{DNAH5 SYNE1 SYNE2 CAPS PIFO}
  \item \texttt{GNLY KLRD1 KLRK1 NKG7 PRF1 GZMB}
  \item \texttt{GNLY NKG7 KLRD1 KLRK1 KLRC1}
  \item \texttt{ATP6V1G3 FOXI1 BSND CLCNKB ASCL3}
  \item \texttt{FOXI1 CFTR ATP6V1G3 BSND RARRES2}
  \item \texttt{PLVAP ACKR1 ERG VWF PECAM1 CDH5}
  \item \texttt{VIM PLVAP ACKR1 MGP PTGDS CXCL14}
  \item \texttt{CPA3 TPSAB1 TPSB2 MS4A2 HDC GATA2}
  \item \texttt{TPSAB1 TPSB2 KIT CPA3 MS4A2}
  \item \texttt{HLA-DPA1 HLA-DRB1 CD74 GPR183 LGALS2}
  \item \texttt{HLA-DPA1 HLA-DPB1 HLA-DRB1 CD80 CD86 CD74}
  \item \texttt{SCGB1A1 SCGB3A1 MUC5AC MUC5B LYPD2 PRR4}
  \item \texttt{SCGB1A1 MUC5AC SCGB3A1 LYPD2}
  \item \texttt{MUC5AC MUC5B LYZ SCGB1A1 SCGB3A1}
  \item \texttt{COL1A2 LUM DCN SFRP2 COL3A1 PDGFRA}
  \item \texttt{PDGFRA COL1A2 COL3A1 VCAN DCN LUM}
  \item \texttt{PDGFRB VIM COL1A2 MGP CXCL14}
  \item \texttt{SST CHGA ASCL1 GRP CALCA SYP}
  \item \texttt{GRP ASCL1 SYT1 CHGA SYP CALCA}
  \item \texttt{HLA-E KLRC1 KLRD1 KLRC2 KLRC3 KLRK1}
  \item \texttt{HLA-E KLRC1 KLRD1 CD8A CD8B}
  \item \texttt{CALR LRP1 GNAI2 FOS JUND MAP2K2}
  \item \texttt{GNAI2 CXCR3 F2R S1PR4 CD69}
  \item \texttt{IFIT1 MX1 OAS2 ISG15 IFITM3 IFIT3}
  \item \texttt{IFIT1 MX1 OAS2 IFIT3 IFI6}
  \item \texttt{KDM1A KMT5A RAD50 ERCC6 ERCC8}
  \item \texttt{TXNIP MAP2K2 ETS1 VEGFA KLF2}
  \item \texttt{IFNG IFNGR1 IFNGR2 CALR LRP1}
  \item \texttt{CCL5 CCR5 CXCL10 CXCR3 F2R}
  \item \texttt{CFTR FOXI1 SCGB1A1 KRT5 FOXJ1 MUC5AC}
\end{enumerate}
\subsection{D5: Healthy Breast Tissue Atlas (\cite{bhat2024single} \textit{et al.})}

\subsubsection{Ontology Queries}

\begin{enumerate}
  \item Luminal hormone sensing cells with estrogen receptor expression in the healthy breast
  \item FOXA1 pioneer transcription factor activity in luminal hormone responsive breast epithelial cells
  \item ER$\alpha$--FOXA1--GATA3 transcription factor network in hormone responsive breast cells
  \item Mature luminal cells with hormone receptor positive identity in breast tissue
  \item Hormone sensing alpha versus beta cell states in breast epithelium
  \item LHS cell-enriched fate factor DACH1 and PI3K pathway regulator INPP4B in breast
  \item Lobular epithelial cells expressing APOD and immunoglobulin genes in breast
  \item Luminal adaptive secretory precursor cells and progenitor identity in breast
  \item ELF5 and EHF transcription factor expression in luminal progenitor breast cells
  \item Alveolar progenitor cell state enriched in Indigenous American breast tissue
  \item BRCA1 associated breast cancer originating from luminal progenitor cells
  \item KIT receptor expression and chromatin accessibility in luminal progenitor cells
  \item MFGE8 and SHANK2 expression in luminal progenitor cells of the breast
  \item LASP basal--luminal intermediate progenitor cell identity in the breast
  \item Basal-myoepithelial cells with TP63 and KRT14 expression in breast
  \item Basal cell chromatin accessibility and TP63 binding site enrichment
  \item Basal alpha and basal beta cell states in breast myoepithelium
  \item SOX10 motif enrichment in basal-myoepithelial cells of the breast
  \item KRT14 KRT17 expression in ductal epithelial and basal cells of breast tissue
  \item Fibroblast heterogeneity and cell states in healthy breast stroma
  \item Genetic ancestry-dependent variability in breast fibroblast cell states
  \item Fibro-prematrix state enrichment in African ancestry breast tissue fibroblasts
  \item PROCR ZEB1 PDGFR$\alpha$ multipotent stromal cells enriched in African ancestry breast
  \item Myofibroblast and inflammatory fibroblast subtypes in breast cancer stroma
  \item SFRP4 and Wnt pathway modulation in breast fibroblasts
  \item Endothelial cell subtypes and vascular markers in breast tissue
  \item Lymphatic endothelial cells expressing LYVE1 in breast stroma
  \item ACKR1 stalk-like endothelial cell subtype in breast vasculature
  \item Vascular endothelial cell heterogeneity in mammary gland microvasculature
  \item Breast tissue angiogenesis and endothelial cell MECOM expression
  \item T lymphocyte markers and immune cell identity in breast tissue
  \item CD4 T cell IL7R expression and chromatin accessibility in breast
  \item CD8 T cell GZMK cytotoxic activity and IFNG signaling in breast tissue
  \item Tissue-resident memory T lymphocyte populations in healthy breast
  \item Adaptive immune surveillance by T cells in mammary gland stroma
  \item Macrophage identity and FCGR3A expression in breast tissue stroma
  \item Macrophage subtypes and tissue-resident immune cells in healthy breast
  \item Breast tissue-resident macrophage phagocytic function and complement expression
  \item Myeloid lineage immune cells and monocyte-derived macrophages in mammary gland
  \item Adipocyte subtypes and lipid metabolism in breast tissue
  \item Adipocyte PLIN1 and FABP4 expression in healthy breast stroma
  \item PLIN1 lipid droplet biology and adipocyte identity in mammary fat pad
  \item Mammary gland adipose tissue and fatty acid binding protein expression
  \item Epithelial cell hierarchy from basal to luminal hormone sensing in breast
  \item CXCL12 chemokine expression in endothelial cells and fibroblasts of breast
  \item VEGFA angiogenic signaling from luminal cells to endothelium in breast
  \item IGF1 paracrine signaling from fibroblasts to luminal cells in breast stroma
  \item Breast tissue microenvironment with stromal and immune cell interactions
  \item Ancestry differences in breast tissue cellular composition and cancer risk
  \item Gene expression differences between ductal and lobular epithelial cells of the breast
\end{enumerate}

\subsubsection{Expression Queries}

\begin{enumerate}
  \item \texttt{FOXA1 ESR1 GATA3 ERBB4 ANKRD30A AFF3 TTC6}
  \item \texttt{MYBPC1 THSD4 CTNND2 DACH1 INPP4B NEK10}
  \item \texttt{ESR1 FOXA1 GATA3 ELOVL5 ANKRD30A}
  \item \texttt{AFF3 TTC6 ERBB4 MYBPC1 THSD4}
  \item \texttt{DACH1 NEK10 CTNND2 INPP4B ELOVL5}
  \item \texttt{APOD IGHA1 IGKC ESR1 FOXA1 GATA3}
  \item \texttt{DUSP1 DPM3 RPL36 IGHA1 IGKC APOD}
  \item \texttt{ELF5 EHF KIT CCL28 KRT15 BARX2 NCALD}
  \item \texttt{MFGE8 SHANK2 SORBS2 AGAP1 ELF5}
  \item \texttt{KRT15 CCL28 KIT INPP4B ELF5}
  \item \texttt{RBMS3 EHF BARX2 NCALD ELF5}
  \item \texttt{ESR1 ELF5 EHF KIT CCL28}
  \item \texttt{ELF5 KIT CCL28 EHF KRT15 BARX2}
  \item \texttt{NCALD BARX2 SHANK2 SORBS2 MFGE8 ELF5}
  \item \texttt{TP63 KRT14 KLHL29 FHOD3 SEMA5A}
  \item \texttt{KLHL13 KLHL29 TP63 KRT14 PTPRT}
  \item \texttt{TP63 KRT14 KRT17 FHOD3 ABLIM3}
  \item \texttt{ST6GALNAC3 PTPRM SEMA5A KLHL29}
  \item \texttt{KRT14 KRT17 TP63 KLHL29 KLHL13 FHOD3}
  \item \texttt{LAMA2 SLIT2 RUNX1T1 COL1A1 COL3A1}
  \item \texttt{COL3A1 POSTN COL1A1 IGF1 ADAM12}
  \item \texttt{CFD MGST1 MFAP5 COL3A1 POSTN}
  \item \texttt{PROCR ZEB1 PDGFRA COL1A1 LAMA2}
  \item \texttt{SFRP4 COL1A1 POSTN LAMA2 SLIT2}
  \item \texttt{COL1A1 PDPN CD34 CXCL12 LAMA2}
  \item \texttt{MECOM LDB2 MMRN1 CXCL12 ACKR1}
  \item \texttt{LYVE1 MECOM LDB2 MMRN1}
  \item \texttt{ACKR1 CXCL12 MECOM LDB2}
  \item \texttt{MECOM LDB2 MMRN1 LYVE1 ACKR1}
  \item \texttt{CXCL12 MECOM LDB2 ACKR1 MMRN1}
  \item \texttt{PTPRC SKAP1 ARHGAP15 THEMIS IL7R}
  \item \texttt{IL7R GZMK PTPRC SKAP1}
  \item \texttt{IFNG GZMK IL7R THEMIS PTPRC}
  \item \texttt{THEMIS ARHGAP15 SKAP1 PTPRC IL7R}
  \item \texttt{PTPRC SKAP1 GZMK IFNG THEMIS ARHGAP15}
  \item \texttt{FCGR3A ALCAM LYVE1 CD163}
  \item \texttt{ALCAM FCGR3A LYVE1 CD14}
  \item \texttt{FCGR3A ALCAM CD163 MERTK}
  \item \texttt{ALCAM LYVE1 FCGR3A CD163 MARCO}
  \item \texttt{PLIN1 FABP4 KIT ADIPOQ LEP}
  \item \texttt{FABP4 PLIN1 ADIPOQ LEP LPL}
  \item \texttt{PLIN1 FABP4 LPL PPARG ADIPOQ}
  \item \texttt{FABP4 PLIN1 KIT ADIPOQ}
  \item \texttt{FOXA1 ELF5 TP63 KRT14 GATA3 ESR1}
  \item \texttt{GATA3 EHF ELF5 FOXA1 KRT15 KRT14 TP63}
  \item \texttt{MECOM PTPRC FCGR3A PLIN1 LAMA2 TP63 FOXA1}
  \item \texttt{CXCL12 LAMA2 MECOM LDB2 COL1A1}
  \item \texttt{ESR1 FOXA1 ELF5 EHF KIT TP63 KRT14}
  \item \texttt{PTPRC FCGR3A FABP4 PLIN1 MECOM}
  \item \texttt{VEGFA LDB2 IGF1 LAMA2 FOXA1 ELF5}
\end{enumerate}

\subsection{D3: Fetal Lung AT2 Organoids (\cite{lim2025novel} \textit{et al.})}

\subsubsection{Ontology Queries}

\begin{enumerate}
  \item Alveolar type 2 cell identity and surfactant protein production in fetal lung organoids
  \item Mature AT2 cell markers and lamellar body formation in fdAT2 organoids
  \item Surfactant protein C maturation and intracellular trafficking in alveolar epithelium
  \item SFTPC processing through endosomal compartments and multivesicular bodies
  \item Surfactant secretion and lamellar body exocytosis in human AT2 cells
  \item ITCH E3 ubiquitin ligase role in SFTPC trafficking and ubiquitination
  \item K63 ubiquitination of surfactant protein C for ESCRT recognition and MVB entry
  \item HECT domain E3 ligase ITCH depletion phenocopying SFTPC-I73T pathogenic variant
  \item Ubiquitome forward genetic screen for SFTPC trafficking effectors
  \item SFTPC relocalisation to plasma membrane and recycling endosomes upon ITCH loss
  \item AT2 stem cell self-renewal and proliferation in fetal lung organoids
  \item FGF7-driven AT2 cell proliferation and surfactant processing balance
  \item Expandable fetal-derived AT2 organoids maintaining identity over passaging
  \item Alveolar type 1 cell differentiation from AT2 organoids via YAP activation
  \item AT2 to AT1 lineage transition through Wnt withdrawal and LATS inhibition
  \item AT1 cell fate markers AQP5 CAV1 AGER in differentiated fdAT2 organoids
  \item CXCL chemokine expressing AT2 subpopulation in fetal lung organoids
  \item Immune response gene expression in alveolar type 2 cells
  \item Chemokine-mediated innate immune signaling in AT2 organoid subsets
  \item Aberrant basal cell differentiation from AT2 cells in organoid culture
  \item Hypoxia-induced airway differentiation of alveolar type 2 cells
  \item Pulmonary neuroendocrine cell differentiation in AT2 organoids
  \item Neuroendocrine progenitor cells co-expressing SFTPC and NE markers
  \item Ciliated cell-like differentiation in fetal AT2 organoid culture
  \item Intermediate transitional cell state between AT2 and differentiated lineages
  \item Surfactant metabolism and lipid transport in fetal alveolar epithelium
  \item Vesicle-mediated transport and lysosome localization in AT2 surfactant processing
  \item Lipid storage membrane transport and vesicle cytoskeleton trafficking in AT2 cells
  \item Wnt signaling pathway maintaining AT2 identity and inhibiting AT1 differentiation
  \item SFTPC-I73T pathogenic variant causing interstitial lung disease and AT2 dysfunction
  \item Toxic gain-of-function effect of misfolded surfactant protein C variants
  \item Transcriptional maturity of fdAT2 organoids compared to adult AT2 and PSC-iAT2
  \item Missing immune response MHC class II genes in fetal versus adult AT2 cells
  \item CRISPRi-mediated depletion of ITCH and UBE2N in fdAT2 organoids
  \item Reversible SFTPC mislocalization after CRISPRi recovery in AT2 organoids
  \item ESCRT complex components HRS VPS28 required for SFTPC MVB entry
  \item Endosomal recycling of SFTPC to plasma membrane upon ubiquitination failure
  \item SUMOylation pathway components UBE2I UBA2 PIAS1 and SFTPC expression regulation
  \item Fetal lung tip progenitor differentiation into mature AT2 cells
  \item EpCAM positive tip epithelial cell isolation and AT2 organoid derivation
  \item SFTPC C-terminal cleavage and proprotein processing in endosomal compartments
  \item proSFTPC plasma membrane transit before endocytosis and maturation
  \item Interstitial lung disease caused by SFTPC variants and AT2 cell dysfunction
  \item Heritable pulmonary fibrosis from SFTPC mistrafficking and toxic accumulation
  \item AT2 medium components dexamethasone cAMP IBMX DAPT for alveolar differentiation
  \item fdAT2 organoid engraftment in mouse precision-cut lung slices and AT1 differentiation
  \item NEDD4-2 HECT domain ligase role in SFTPC ubiquitination and maturation
  \item Cell type heterogeneity and proportions across fdAT2 organoid lines
  \item fdAT2 organoid stability over long-term passaging and cryopreservation
  \item Genetic manipulation of fetal AT2 organoids using lentiviral CRISPRi system
\end{enumerate}

\subsubsection{Expression Queries}

\begin{enumerate}
  \item \texttt{SFTPC SFTPB SFTPA1 SFTPA2 NAPSA LAMP3}
  \item \texttt{SFTPC SFTPB ABCA3 LAMP3 HOPX NKX2-1}
  \item \texttt{NKX2-1 SLC34A2 LPCAT1 HOPX CEACAM6}
  \item \texttt{SFTPC SFTPD SFTA3 CD36 CAV1 SLC34A2}
  \item \texttt{SFTPA1 SFTPA2 SFTPB SFTPC SFTPD}
  \item \texttt{ITCH UBE2N HRS VPS28 RABGEF1 EEA1}
  \item \texttt{ITCH NEDD4 NEDD4L UBE2N UBE2I}
  \item \texttt{EEA1 MICALL1 LAMP3 HRS VPS28}
  \item \texttt{UBE2I UBA2 PIAS1 ITCH RABGEF1}
  \item \texttt{ABCA3 LAMP3 NAPSA CKAP4 ZDHHC2 CTSH}
  \item \texttt{ABCA3 SFTPB SFTPC LAMP3 P2RY2 LMCD1}
  \item \texttt{MKI67 PCNA TOP2A SFTPC NKX2-1}
  \item \texttt{MKI67 PCNA CDK1 CCNB1 SFTPC}
  \item \texttt{CXCL1 CXCL2 CXCL3 CCL2 SFTPC}
  \item \texttt{CXCL1 CXCL3 CCL2 CCL4 CCL4L1}
  \item \texttt{CXCL1 CXCL2 HLA-DPA1 HLA-DPB1 CCL2}
  \item \texttt{HLA-DQB1 HLA-DMA HLA-DMB HLA-DRA HLA-DOA}
  \item \texttt{HLA-DPA1 HLA-DPB1 HLA-DRA CD86 TNF}
  \item \texttt{AQP5 CAV1 AGER HOPX}
  \item \texttt{CAV1 AGER AQP5 PDPN}
  \item \texttt{TP63 KRT5 KRT14 KRT15 SOX2}
  \item \texttt{KRT5 KRT14 TP63 LAMB3 COL17A1}
  \item \texttt{ASCL1 NEUROD1 GRP CHGA SYP CALCA}
  \item \texttt{GRP ASCL1 SYT1 CHGA SYP}
  \item \texttt{ASCL1 GRP SFTPC NKX2-1}
  \item \texttt{FOXJ1 DNAH5 CAPS PIFO RSPH1}
  \item \texttt{FOXJ1 DNAH5 DNAI1 RSPH1 CAPS}
  \item \texttt{SOX2 SOX9 NKX2-1 SFTPC TP63}
  \item \texttt{SOX2 NKX2-1 HOPX CAV1}
  \item \texttt{CTNNB1 TCF7L2 AXIN2 WNT3A LGR5}
  \item \texttt{SFTPC NKX2-1 HOPX SFTPB ABCA3 MKI67}
  \item \texttt{NAPSA ABCA3 SFTA3 SFTPD LAMP3 HOPX}
  \item \texttt{SFTPC ITCH EEA1 LAMP3 MICALL1 ABCA3}
  \item \texttt{SFTPC NAPSA CTSH LAMP3 ITCH UBE2N}
  \item \texttt{SFTPC CXCL1 CXCL2 NKX2-1 LAMP3}
  \item \texttt{CDH1 TJP1 EPCAM SFTPC NKX2-1}
  \item \texttt{ITCH HRS VPS28 UBE2N RABGEF1 PIAS1 UBE2I UBA2}
  \item \texttt{ITCH NEDD4 NEDD4L HRS UBAP1 USP8}
  \item \texttt{MKI67 TOP2A PCNA CDK1 CCNB1 CCNA2}
  \item \texttt{SFTPC TP63 ASCL1 FOXJ1 NKX2-1}
  \item \texttt{SFTPC SFTPB ASCL1 GRP TP63 KRT5}
  \item \texttt{SFTPC CAV1 AGER AQP5 HOPX NKX2-1}
  \item \texttt{LAMP3 ABCA3 SFTPB SFTPC NAPSA CD36}
  \item \texttt{CKAP4 ZDHHC2 SLC34A2 CTSH SFTPC}
  \item \texttt{CXCL1 CXCL2 CXCL3 CCL2 CCL4 TNF}
  \item \texttt{SOX9 NKX2-1 SFTPC SFTPB LAMP3}
  \item \texttt{SFTPC NKX2-1 ASCL1 NEUROD1 GRP MKI67}
  \item \texttt{SFTA3 SFTPD NAPSA NKX2-1 CKAP4 ZDHHC2 SLC34A2 CTSH SFTPA1 SFTPA2 SFTPC SFTPB}
  \item \texttt{ITCH SFTPC LAMP3 ABCA3 UBE2N NAPSA}
  \item \texttt{SFTPC CXCL1 MKI67 TP63 ASCL1 FOXJ1 SOX2 CAV1}
\end{enumerate}

\subsection{D2: High-Risk Neuroblastoma (\cite{yu2025longitudinal} \textit{et al.})}

\subsubsection{Ontology Queries}

\begin{enumerate}
  \item Neuroblast neoplastic cell of sympathetic nervous system expressing PHOX2B and ISL1
  \item Neuroblastoma tumor cell with MYCN amplification and proliferative phenotype
  \item Adrenergic neuroblast expressing catecholamine biosynthesis enzymes tyrosine hydroxylase
  \item Neuroblastoma cell with calcium and synaptic signaling pathway enrichment
  \item Dopaminergic neuroblast expressing dopamine transporter and metabolic genes
  \item Proliferating neuroblastoma cell with cell cycle and DNA replication markers
  \item Mesenchymal neuroblastoma cell state expressing extracellular matrix genes and YAP1
  \item Intermediate OXPHOS neuroblast with ribosomal gene expression and oxidative phosphorylation
  \item EZH2 expressing neuroblastoma cell PRC2 polycomb repressive complex chromatin regulation
  \item Neuroblastoma cell ERBB4 receptor expressing epidermal growth factor signaling
  \item Neuroblast with adrenergic transcription factor PHOX2A PHOX2B GATA3 expression
  \item Neural crest derived neoplastic cell in pediatric tumor expressing chromogranin
  \item Neuroblastoma cell immune evasion NECTIN2 and checkpoint ligand expression
  \item Mesenchymal transition state in neuroblastoma with AP-1 transcription factors
  \item Tumor associated macrophage in neuroblastoma microenvironment CD68 CD163 expressing
  \item Pro-inflammatory macrophage IL18 expressing anti-tumor immune response
  \item Pro-angiogenic macrophage VCAN expressing promoting tumor vascularization
  \item Immunosuppressive macrophage C1QC SPP1 complement expressing in tumor
  \item Tissue resident macrophage F13A1 expressing phagocytic function in neuroblastoma
  \item Lipid associated macrophage HS3ST2 with metabolic phenotype in tumor
  \item Macrophage secreting HB-EGF ligand for ERBB4 receptor activation on neuroblasts
  \item CCL4 expressing pro-angiogenic macrophage chemokine signaling in tumor
  \item Proliferating macrophage MKI67 TOP2A expanding after chemotherapy
  \item THY1 positive macrophage undefined myeloid phenotype in neuroblastoma
  \item T cell lymphocyte infiltrating neuroblastoma tumor expressing CD247 CD96
  \item Cytotoxic T cell with granzyme perforin mediated tumor cell killing
  \item Tumor infiltrating T lymphocyte immune response to neuroblastoma
  \item B cell lymphocyte PAX5 MS4A1 in neuroblastoma tumor immune microenvironment
  \item B lymphocyte humoral immunity and antigen presentation in pediatric tumor
  \item Dendritic cell IRF8 FLT3 antigen presentation priming T cell responses in tumor
  \item Professional antigen presenting dendritic cell MHC class II expression
  \item Fibroblast stromal cell PDGFRB DCN extracellular matrix production in neuroblastoma
  \item Cancer associated fibroblast FAP ACTA2 expressing in tumor stroma
  \item Neural crest derived endoneurial fibroblast in neuroblastoma tissue
  \item Schwann cell PLP1 CDH19 myelinating glial cell in neuroblastoma microenvironment
  \item Schwann cell precursor neural crest lineage expanding after therapy
  \item Endothelial cell PECAM1 PTPRB vascular marker in neuroblastoma tumor vasculature
  \item Tumor endothelium blood vessel lining cell expressing vascular endothelial markers
  \item Adrenal cortex cell steroidogenesis CYP11A1 CYP11B1 adjacent normal tissue
  \item Cortical cell of adrenal gland steroid hormone biosynthesis normal adjacent tissue
  \item Hepatocyte ALB expressing liver cell from adjacent normal tissue in neuroblastoma biopsy
  \item Kidney cell renal tissue PKHD1 from adjacent normal tissue in neuroblastoma specimen
  \item Chemotherapy induced tumor microenvironment rewiring macrophage expansion after therapy
  \item HB-EGF ERBB4 paracrine signaling axis between macrophage and neuroblast promoting ERK
  \item Tumor immune evasion and antigen presentation in neuroblastoma
  \item VEGFA angiogenesis signaling in neuroblastoma tumor microenvironment
  \item Immune cell infiltration in high-risk neuroblastoma T cell B cell macrophage
  \item THBS1 CD47 don't eat me signal between macrophage and neuroblastoma cell
  \item Neuroblastoma cell expressing ALK receptor tyrosine kinase oncogenic driver
  \item Tumor microenvironment cell diversity neuroblasts fibroblasts Schwann endothelial macrophages
\end{enumerate}

\subsubsection{Expression Queries}

\begin{enumerate}[label=Q\arabic*., start=51]
  \item \texttt{PHOX2B ISL1 HAND2 TH DBH DDC CHGA}
  \item \texttt{MYCN MKI67 TOP2A EZH2 SMC4 BIRC5}
  \item \texttt{PHOX2A PHOX2B GATA3 ASCL1 ISL1 HAND2}
  \item \texttt{CACNA1B SYN2 KCNMA1 KCNQ3 GPC5 CREB5}
  \item \texttt{SLC18A2 TH DDC AGTR2 ATP2A2 PHOX2B}
  \item \texttt{MKI67 TOP2A EZH2 SMC4 BIRC5 BUB1B ASPM KIF11}
  \item \texttt{YAP1 FN1 VIM COL1A1 SERPINE1 SPARC THBS2}
  \item \texttt{ERBB4 EGFR HBEGF TGFA EREG AREG}
  \item \texttt{NECTIN2 CD274 B2M HLA-A HLA-B PHOX2B}
  \item \texttt{JUN FOS JUNB JUND FOSL2 BACH1 BACH2}
  \item \texttt{CHGA CHGB PHOX2B ISL1 NTRK1 RET}
  \item \texttt{ETS1 ETV6 ELF1 KLF6 KLF7 RUNX1 ZNF148}
  \item \texttt{ALK MYCN NTRK2 PHOX2B TH}
  \item \texttt{CD68 CD163 CD86 CSF1R MRC1 SPP1}
  \item \texttt{IL18 CD68 CD163 CD86 HLA-DRA CSF1R}
  \item \texttt{VCAN VEGFA CD68 CD163 SPP1 EGFR}
  \item \texttt{C1QC SPP1 CD68 CD163 APOE TREM2}
  \item \texttt{F13A1 CD68 CD163 MRC1 LYVE1 CSF1R}
  \item \texttt{HS3ST2 CYP27A1 CD68 CD163 APOE LPL}
  \item \texttt{HBEGF TGFA EREG AREG CD68 CD163}
  \item \texttt{CCL4 CD68 CD163 VEGFA CSF1R CCL3}
  \item \texttt{THY1 CD68 CD163 MRC1 CSF1R CD86}
  \item \texttt{CD247 CD96 CD3D CD3E CD8A CD4}
  \item \texttt{GZMA GZMB PRF1 IFNG CD8A CD3D}
  \item \texttt{PAX5 MS4A1 CD19 CD79A HLA-DRA HLA-DRB1}
  \item \texttt{IRF8 FLT3 CLEC9A CD1C CD80 HLA-DRA}
  \item \texttt{PDGFRB DCN LUM COL1A1 COL1A2 VIM}
  \item \texttt{FAP ACTA2 COL1A1 PDGFRA DCN LUM}
  \item \texttt{PLP1 CDH19 SOX10 MPZ MBP S100B}
  \item \texttt{PECAM1 PTPRB CDH5 VWF KDR FLT1}
  \item \texttt{CYP11A1 CYP11B1 CYP17A1 STAR NR5A1}
  \item \texttt{ALB DCDC2 HNF4A APOB}
  \item \texttt{PKHD1 PAX2 WT1 SLC12A1}
  \item \texttt{PHOX2B CD68 CD3D MS4A1 PECAM1 DCN PLP1}
  \item \texttt{HBEGF ERBB4 CD68 PHOX2B MAPK1}
  \item \texttt{VCAN THBS1 CD47 ITGB1 CD68 PHOX2B}
  \item \texttt{HLA-A HLA-B HLA-C B2M HLA-DRA HLA-DRB1}
  \item \texttt{VEGFA KDR FLT1 NRP1 GPC1 PECAM1}
  \item \texttt{CD68 IL18 VCAN C1QC SPP1 F13A1 HS3ST2 CCL4 THY1}
  \item \texttt{PHOX2B MKI67 TOP2A YAP1 CACNA1B SLC18A2}
  \item \texttt{APOE LDLR VLDLR LPL HS3ST2 CD68}
  \item \texttt{THBS1 ITGB1 ITGA3 LRP5 CD47 FN1}
  \item \texttt{COL1A1 COL1A2 COL4A1 COL4A2 FN1 VIM SPARC}
  \item \texttt{MAPK1 MAPK3 AKT1 ERBB4 EGFR HBEGF}
  \item \texttt{CD274 PDCD1 CTLA4 TIGIT LAG3 NECTIN2}
  \item \texttt{PHOX2B CD68 PLP1 PECAM1 DCN IRF8 PAX5 CD247}
  \item \texttt{CYP11A1 ALB PKHD1 PHOX2B CD68}
  \item \texttt{PHOX2B HBEGF ERBB4 VCAN SPP1 CD163 VEGFA}
  \item \texttt{MKI67 TOP2A PCNA CDK1 CCNB1 EZH2 MELK}
  \item \texttt{PHOX2B ISL1 CD68 CD163 CD3D MS4A1 PLP1 PECAM1 DCN CYP11A1 ALB}
\end{enumerate}


\subsection{D3: Immune Checkpoint Blockade Multi-Cancer (\cite{gondal2025integrated} \textit{et al.})}

\subsubsection{Ontology Queries}

\begin{enumerate}
  \item Malignant cancer cell expressing immune checkpoint ligand PD-L1 for immune evasion
  \item Tumor cell immune evasion through HLA downregulation and B2M loss
  \item Melanoma cancer cell expressing MITF MLANA PMEL lineage markers
  \item Breast cancer epithelial cell markers EPCAM KRT8 KRT18 KRT19 in ICB treated tumors
  \item Tumor cell proliferation and cell cycle markers in malignant cells
  \item Cancer cell VEGFA and TGFB1 immunosuppressive signaling in tumor microenvironment
  \item Epithelial mesenchymal transition EMT markers in cancer cells during ICB treatment
  \item Effector CD8 T cell cytotoxic function with granzyme and perforin expression
  \item Activated CD8 T cell expressing IFNG and TNF anti-tumor cytokines
  \item CD8 T cell exhaustion with PD-1 LAG3 TIM3 TIGIT checkpoint receptor co-expression
  \item TOX transcription factor driving T cell exhaustion program in chronic antigen stimulation
  \item Central memory CD8 T cell with TCF7 and IL7R expression for long-lived immunity
  \item Naive CD8 T cell expressing CCR7 SELL before antigen encounter
  \item CD8-positive T cell co-stimulatory receptor 4-1BB ICOS upon activation
  \item CD4 positive helper T cell TCR signaling and cytokine production
  \item Regulatory T cell FOXP3 expressing immunosuppressive function in tumor
  \item T follicular helper cell CXCR5 BCL6 supporting B cell responses in tertiary lymphoid structures
  \item Th17 helper T cell IL17A RORC inflammatory response in tumor microenvironment
  \item CD8-positive CD28-negative regulatory T cell with suppressive function
  \item Natural killer T cell NKT innate cytotoxicity with KLRD1 and NKG7 expression
  \item NK cell mediated tumor killing through NCR1 and KLRB1 receptor activation
  \item B cell CD19 MS4A1 CD79A antigen presentation and humoral immunity in tumor
  \item Plasma cell antibody secreting immunoglobulin production SDC1 MZB1
  \item Tertiary lymphoid structure B cell and plasma cell formation in ICB-responsive tumors
  \item Tumor associated macrophage M2 polarization CD163 MRC1 immunosuppressive function
  \item Macrophage complement expression C1QA C1QB and TREM2 in tumor microenvironment
  \item Classical monocyte CD14 LYZ infiltration into tumor during checkpoint blockade
  \item Dendritic cell antigen presentation CD80 CD86 priming T cell responses
  \item Plasmacytoid dendritic cell IRF7 LILRA4 type I interferon production
  \item Myeloid cell general CSF1R ITGAM expressing innate immune population
  \item Mast cell KIT TPSB2 CPA3 in allergic and inflammatory tumor responses
  \item Microglial cell brain resident macrophage in melanoma brain metastasis
  \item Cancer associated fibroblast FAP ACTA2 COL1A1 producing extracellular matrix
  \item Myofibroblast ACTA2 TAGLN contractile smooth muscle actin expression in tumor stroma
  \item Tumor endothelial cell PECAM1 CDH5 VWF vascular marker expression
  \item Melanocyte pigmentation pathway MITF TYR TYRP1 DCT lineage genes
  \item Hematopoietic multipotent progenitor cell stem cell marker expression
  \item PD-1 blockade restoring effector CD8 T cell anti-tumor cytotoxicity
  \item CTLA-4 blockade enhancing CD4 helper T cell and reducing Treg suppression
  \item T cell clonal replacement and expansion following PD-1 checkpoint inhibition
  \item TCF4 dependent resistance program in mesenchymal-like melanoma cells
  \item T cell exclusion program in tumor cells resisting checkpoint blockade therapy
  \item Antigen processing and MHC class I presentation in tumor cells
  \item MHC class II antigen presentation by professional antigen presenting cells
  \item Interferon gamma response driving PD-L1 upregulation on tumor cells
  \item Tumor infiltrating lymphocyte diversity including T B and NK cells
  \item Liver cancer hepatocellular carcinoma markers ALB AFP GPC3 in ICB dataset
  \item Clear cell renal carcinoma CA9 PAX8 markers in kidney cancer patients
  \item Basal cell carcinoma Hedgehog pathway PTCH1 GLI1 GLI2 SHH signaling
  \item Lymphocyte general population in tumor immune microenvironment
\end{enumerate}

\subsubsection{Expression Queries}

\begin{enumerate}
  \item \texttt{CD274 PDCD1LG2 B2M HLA-A CD47 IDO1 VEGFA}
  \item \texttt{MITF MLANA PMEL TYR DCT SOX10 TYRP1}
  \item \texttt{EPCAM KRT8 KRT18 KRT19 MUC1 CDH1 ESR1}
  \item \texttt{MKI67 TOP2A PCNA CD274 B2M TGFB1}
  \item \texttt{PRF1 GZMA GZMB GZMK GNLY NKG7 IFNG}
  \item \texttt{GZMB PRF1 IFNG TNF FASLG NKG7 CD8A}
  \item \texttt{CD69 ICOS TNFRSF9 IFNG GZMB CD8A}
  \item \texttt{PDCD1 LAG3 HAVCR2 TIGIT TOX ENTPD1}
  \item \texttt{TOX TOX2 PDCD1 HAVCR2 LAG3 TIGIT BTLA}
  \item \texttt{TCF7 LEF1 CCR7 SELL IL7R CD8A CD8B}
  \item \texttt{CCR7 SELL TCF7 LEF1 IL7R CD3D}
  \item \texttt{CD4 CD3D CD3E IL7R CD28 ICOS TCF7}
  \item \texttt{FOXP3 IL2RA CTLA4 IKZF2 TNFRSF18 TIGIT}
  \item \texttt{CXCR5 BCL6 ICOS PDCD1 CD4 CD3D}
  \item \texttt{RORC IL17A IL23R CCR6 CD4 CD3E}
  \item \texttt{CD8A GZMB PRF1 LAG3 CTLA4 PDCD1}
  \item \texttt{KLRD1 KLRK1 NKG7 GNLY PRF1 GZMB NCAM1}
  \item \texttt{NCAM1 NCR1 KLRB1 KLRC1 GZMB IFNG}
  \item \texttt{CD19 MS4A1 CD79A CD79B HLA-DRA HLA-DRB1}
  \item \texttt{SDC1 MZB1 JCHAIN IGHG1 IGKC CD79A}
  \item \texttt{CD163 MRC1 MSR1 MARCO CD68 APOE TREM2}
  \item \texttt{C1QA C1QB APOE TREM2 CD68 SPP1}
  \item \texttt{CD14 FCGR3A S100A8 S100A9 LYZ CSF1R}
  \item \texttt{CD80 CD86 CD83 CCR7 HLA-DRA CLEC9A}
  \item \texttt{LILRA4 IRF7 IRF8 IL3RA NRP1}
  \item \texttt{ITGAM CSF1R CD68 LYZ S100A8 S100A9}
  \item \texttt{KIT TPSB2 TPSAB1 CPA3 HPGDS HDC}
  \item \texttt{P2RY12 TMEM119 CX3CR1 CSF1R AIF1}
  \item \texttt{FAP ACTA2 COL1A1 COL1A2 PDGFRA DCN LUM}
  \item \texttt{ACTA2 TAGLN MYH11 COL1A1 PDGFRB VIM}
  \item \texttt{PECAM1 CDH5 VWF KDR FLT1 ENG}
  \item \texttt{MITF TYR TYRP1 DCT MLANA PMEL SOX10}
  \item \texttt{CD34 KIT FLT3 PROM1 THY1 PTPRC}
  \item \texttt{CD3D CD3E CD8A CD4 TRAC TRBC1}
  \item \texttt{HLA-DRA HLA-DRB1 HLA-DPA1 HLA-DPB1 CD74 CIITA}
  \item \texttt{HLA-A HLA-B HLA-C B2M TAP1 TAP2}
  \item \texttt{PDCD1 CD274 CTLA4 CD80 CD86 LAG3 HAVCR2}
  \item \texttt{CD274 CD47 IDO1 GZMB PRF1 IFNG}
  \item \texttt{CD8A CD4 MS4A1 CD68 PECAM1 FAP EPCAM NCAM1}
  \item \texttt{GZMB IFNG FOXP3 CD163 CD274 MS4A1 PECAM1}
  \item \texttt{ALB AFP GPC3 EPCAM KRT19}
  \item \texttt{CA9 PAX8 MME EPCAM VEGFA}
  \item \texttt{PTCH1 GLI1 GLI2 EPCAM KRT14}
  \item \texttt{ERBB2 ESR1 EPCAM KRT8 KRT18 MUC1}
  \item \texttt{CCR7 SELL TCF7 PDCD1 TOX GZMB PRF1}
  \item \texttt{IFNG CD274 STAT1 IRF1 B2M HLA-A}
  \item \texttt{CD8A CD4 FOXP3 CXCR5 RORC CCR7 KLRD1 CD3D}
  \item \texttt{CD68 CD163 CD14 S100A8 CD80 KIT LILRA4 ITGAM}
  \item \texttt{FAP ACTA2 PECAM1 CDH5 COL1A1 PDGFRA VWF}
  \item \texttt{CD274 GZMB CD68 MS4A1 FAP PECAM1 MITF FOXP3 CD8A KIT LILRA4}
\end{enumerate}


\subsection{D6: First-Trimester Human Brain (\cite{mannens2025chromatin} \textit{et al.})}

\subsubsection{Ontology Queries}

\begin{enumerate}
  \item GABAergic inhibitory neuron differentiation in developing human midbrain
  \item Midbrain GABAergic neuron OTX2 GATA2 TAL2 transcription factor expression
  \item Cortical interneuron derived from medial ganglionic eminence LHX6 DLX2
  \item Interneuron diversity parvalbumin somatostatin VIP subtypes developing cortex
  \item TAL2 expressing midbrain GABAergic neurons linked to major depressive disorder
  \item Lateral and caudal ganglionic eminence interneuron migration in telencephalon
  \item Medial ganglionic eminence derived parvalbumin somatostatin interneuron
  \item SOX14 expressing midbrain GABAergic neuron thalamic migration
  \item Glutamatergic excitatory neuron in developing human telencephalon cortex
  \item Telencephalic glutamatergic neuron LHX2 BHLHE22 cortical layer specification
  \item Hindbrain glutamatergic neuron ATOH1 MEIS1 cerebellar granule cell
  \item Deep layer cortical neuron FEZF2 BCL11B corticospinal projection
  \item SATB2 expressing telencephalic excitatory neuron callosal projection
  \item Upper layer cortical neuron CUX1 CUX2 RORB intracortical connectivity
  \item EMX2 transcription factor dorsal telencephalon glutamatergic identity
  \item Purkinje cell differentiation in developing cerebellum PTF1A ESRRB lineage
  \item Purkinje neuron ESRRB oestrogen-related nuclear receptor cerebellum specific
  \item Cerebellar Purkinje progenitor PTF1A ASCL1 NEUROG2 ventricular zone
  \item TFAP2B LHX5 activation of ESRRB enhancer in Purkinje neuroblast
  \item RORA FOXP2 EBF3 late Purkinje maturation gene regulatory network
  \item Cerebellar granule neuron ATOH1 MEIS1 external granular layer
  \item Radial glial cell neural stem cell SOX2 PAX6 NES in developing brain
  \item Radial glia to glioblast transition NFI factor maturation NFIA NFIB NFIX
  \item Neural progenitor cell proliferation and neurogenesis in ventricular zone
  \item Loss of stemness and glial fate restriction by NFI transcription factors
  \item Progenitor cell dividing in developing human brain VIM HES1 proliferating
  \item Notch signaling DLL1 JAG1 NOTCH1 lateral inhibition neurogenesis
  \item Glioblast astrocyte precursor GFAP S100B AQP4 BCAN TNC fetal brain
  \item Astrocyte maturation and glial scar markers in developing brain
  \item Oligodendrocyte precursor cell OLIG2 PDGFRA SOX10 specification
  \item Oligodendrocyte differentiation MBP MOG PLP1 myelination fetal brain
  \item Committed oligodendrocyte precursor SOX10 lineage commitment
  \item Dopaminergic neuron midbrain TH NR4A2 substantia nigra ventral tegmental area
  \item Serotonergic neuron raphe nucleus TPH2 SLC6A4 FEV brainstem
  \item FOXA2 LMX1A floor plate derived dopaminergic neuron specification
  \item Endothelial cell blood--brain barrier CLDN5 PECAM1 CDH5 fetal brain
  \item Pericyte PDGFRB RGS5 FOXF2 cerebral vasculature developing brain
  \item Vascular leptomeningeal cell FOXC1 meningeal fibroblast DCN COL1A1
  \item Vascular smooth muscle cell ACTA2 MYH11 cerebral artery
  \item Microglial cell CX3CR1 P2RY12 TMEM119 brain resident macrophage
  \item Border-associated macrophage RUNX1 haematopoietic origin fetal brain
  \item Immature T cell and leukocyte infiltration in developing fetal brain
  \item Schwann cell MPZ CDH19 SOX10 neural crest derived myelinating peripheral glial
  \item Sensory neuron dorsal root ganglion NTRK1 ISL1 peripheral nervous system
  \item Glycinergic neuron SLC6A5 GLRA1 inhibitory spinal cord hindbrain
  \item Neuroblast immature migrating neuron fetal cortex RBFOX3 NEFM
  \item Major depressive disorder MDD midbrain GABAergic neuron NEGR1 LRFN5
  \item Schizophrenia cortical interneuron medial ganglionic eminence SATB2
  \item Attention deficit hyperactivity disorder ADHD cerebellar Purkinje
  \item Autism spectrum disorder hindbrain neuroblast brainstem involvement
\end{enumerate}

\subsubsection{Expression Queries}

\begin{enumerate}
  \item \texttt{GAD1 GAD2 SLC32A1 DLX2 DLX5 LHX6}
  \item \texttt{OTX2 GATA2 TAL2 SOX14 GAD2 SLC32A1}
  \item \texttt{PVALB SST VIP LAMP5 SNCG ADARB2}
  \item \texttt{DLX1 DLX2 DLX5 DLX6 MEIS2 LHX6}
  \item \texttt{GAD1 GAD2 SLC32A1 TFAP2B OTX2}
  \item \texttt{TAL2 SOX14 GAD2 OTX2 GATA2}
  \item \texttt{SLC17A7 SLC17A6 SATB2 TBR1 FEZF2 BCL11B}
  \item \texttt{EMX2 LHX2 BHLHE22 CUX1 CUX2 RORB}
  \item \texttt{ATOH1 MEIS1 MEIS2 SLC17A6 RBFOX3}
  \item \texttt{FEZF2 BCL11B TBR1 SATB2 SLC17A7}
  \item \texttt{CUX1 CUX2 RORB LHX2 BHLHE22 EMX2}
  \item \texttt{PTF1A ASCL1 NEUROG2 NHLH1 NHLH2 TFAP2B}
  \item \texttt{ESRRB RORA PCP4 FOXP2 EBF3 LHX5}
  \item \texttt{LHX5 LHX1 PAX2 TFAP2B DMBX1 NHLH2}
  \item \texttt{ESRRB PCP4 RORA EBF1 EBF3 FOXP2 LHX1}
  \item \texttt{SOX2 PAX6 NES VIM HES1 HES5 FABP7}
  \item \texttt{NFIA NFIB NFIX SOX9 FABP7}
  \item \texttt{SOX2 HES1 HES5 PAX6 NES VIM}
  \item \texttt{NOTCH1 NOTCH2 DLL1 JAG1 HES1 HES5}
  \item \texttt{GFAP S100B AQP4 ALDH1L1 BCAN TNC}
  \item \texttt{OLIG1 OLIG2 SOX10 PDGFRA CSPG4}
  \item \texttt{MBP MOG PLP1 MAG SOX10}
  \item \texttt{OLIG2 SOX10 PDGFRA NKX2-2 OLIG1}
  \item \texttt{TH DDC SLC6A3 SLC18A2 NR4A2 LMX1A FOXA2}
  \item \texttt{FOXA2 LMX1A NR4A2 TH DDC SLC18A2}
  \item \texttt{TPH2 SLC6A4 FEV DDC SLC18A2}
  \item \texttt{SLC6A5 GLRA1 SLC32A1 GAD1}
  \item \texttt{RBFOX3 SNAP25 SYT1 NEFM NEFL TUBB3}
  \item \texttt{NEFM NEFL MAP2 TUBB3 SYT1}
  \item \texttt{CLDN5 PECAM1 CDH5 ERG FLT1 VWF}
  \item \texttt{PDGFRB RGS5 ACTA2 MYH11 COL1A2}
  \item \texttt{ACTA2 MYH11 PDGFRB TAGLN}
  \item \texttt{DCN LUM COL1A1 COL1A2 FOXC1 COL3A1}
  \item \texttt{FOXC1 FOXF2 DCN COL1A2 LUM}
  \item \texttt{AIF1 CX3CR1 P2RY12 TMEM119 HEXB CSF1R}
  \item \texttt{RUNX1 SPI1 CSF1R AIF1 CD68}
  \item \texttt{AIF1 HEXB P2RY12 TMEM119 CX3CR1}
  \item \texttt{CD3D CD3E CD3G PTPRC CD2}
  \item \texttt{MPZ CDH19 SOX10 MBP PLP1}
  \item \texttt{NTRK1 NTRK2 ISL1 PRPH SNAP25}
  \item \texttt{RBFOX3 SLC17A6 GAD2 NEFM SNAP25}
  \item \texttt{NEFM NEFL RBFOX3 TUBB3 DCX}
  \item \texttt{NEGR1 BTN3A2 LRFN5 SCN8A RGS6 MYCN}
  \item \texttt{OTX2 GATA2 MEIS2 PRDM10 MYCN}
  \item \texttt{CTCF MECP2 YY1 RAD21 SMC3}
  \item \texttt{SHH PTCH1 GLI1 GLI2 FOXA2 NKX2-1}
  \item \texttt{WNT5A CTNNB1 LEF1 TCF7L2 AXIN2}
  \item \texttt{BMP4 BMPR1A SMAD1 ID1 ID3}
  \item \texttt{VEGFA KDR FLT1 PDGFB PDGFRB CLDN5}
  \item \texttt{SOX2 PAX6 OLIG2 GFAP RBFOX3 GAD2 SLC17A7}
\end{enumerate}
\clearpage
\section{Example of plot on Cystic Fibrosis Dataset}
\begin{figure*}[ht]
\centering
\makebox[\textwidth][c]{%
  \includegraphics[width=0.7\textwidth]{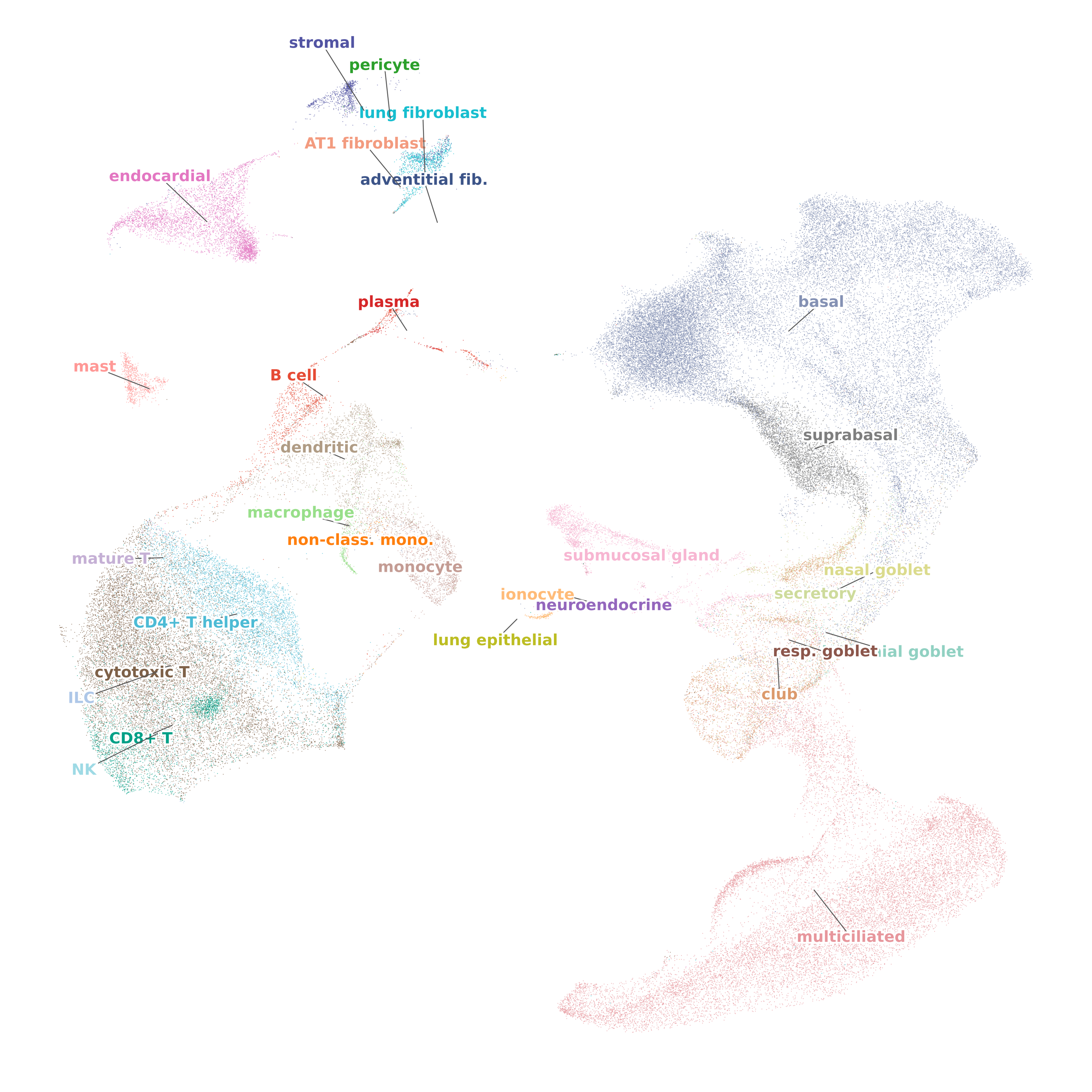}%
}
\caption{Cell-level UMAP of the cystic fibrosis airway dataset (D1) colored by Cell Ontology annotation. Approximately 96,000 cells are shown across 30 annotated cell types spanning immune (T~cells, B~cells, NK~cells, macrophages, monocytes, dendritic cells, mast cells), epithelial (basal, suprabasal, multiciliated, secretory, goblet, club, ionocyte, neuroendocrine), and stromal (fibroblasts, pericytes, endocardial cells) compartments. Labels are placed at cluster centroids with iterative repulsion to minimize overlap.}
\label{fig:architecture}
\end{figure*}
\begin{figure*}[ht]
\centering
\makebox[\textwidth][c]{%
  \includegraphics[width=0.8\textwidth]{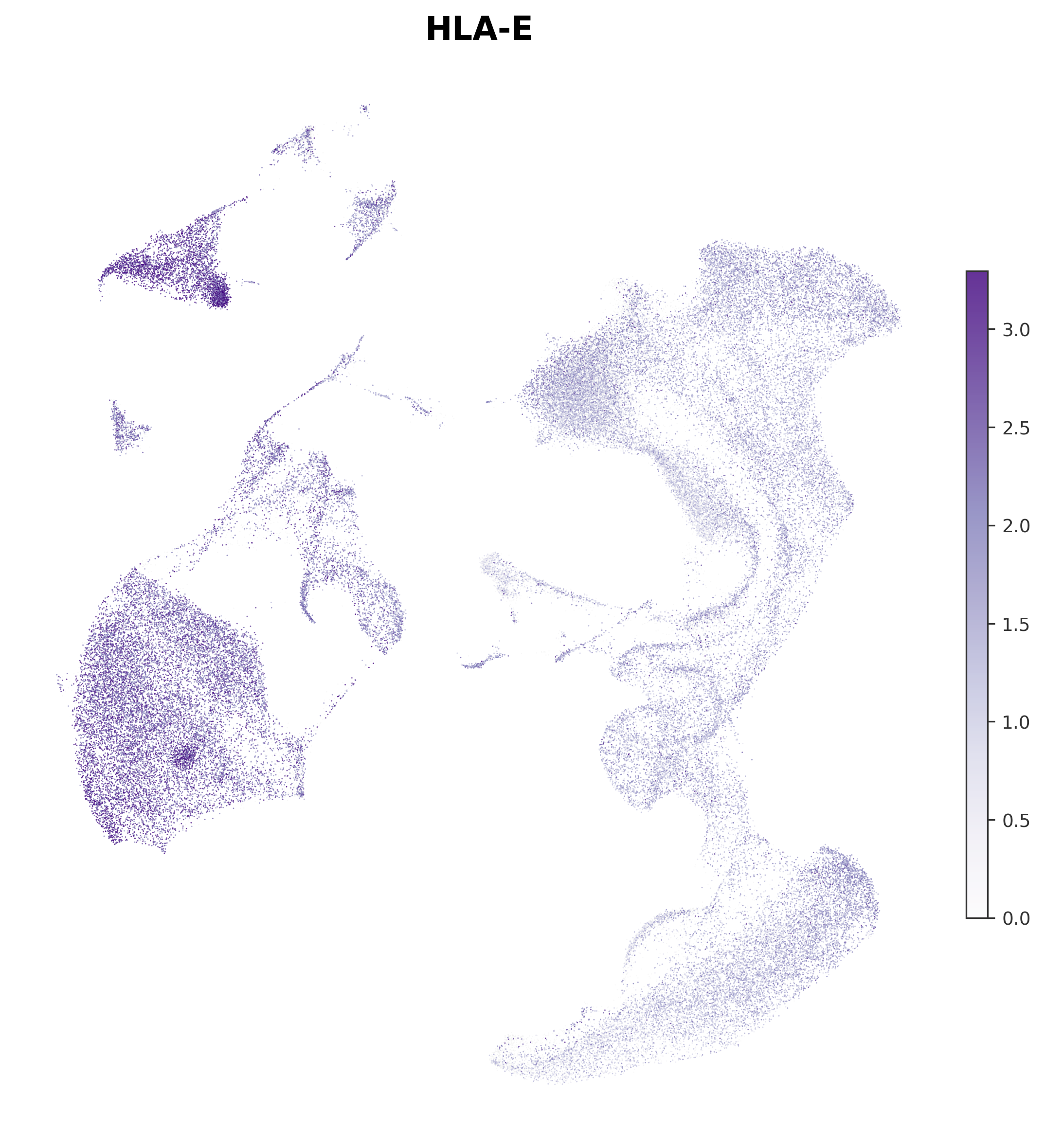}%
}
\caption{Expression of \textit{HLA-E} projected onto the cell-level UMAP of the cystic fibrosis airway dataset (D1). Color intensity (purple gradient) indicates normalized expression level, with non-expressing cells shown in grey. \textit{HLA-E} is most highly expressed in immune cell clusters, particularly CD8$^+$ T~cells and NK~cells, consistent with its role as a ligand for the NKG2A inhibitory receptor. Moderate expression is observed across epithelial populations including basal cells, supporting the HLA-E/NKG2A immune checkpoint axis identified by Berg \textit{et al.}}
\label{fig:architecture}
\end{figure*}
\newpage

\end{appendix}
\end{document}